\documentclass[aps,prb,showpacs,twocolumn,superscriptaddress]{revtex4-1}

\usepackage{amsmath}
\usepackage{amsfonts}
\usepackage{amssymb}
\usepackage{dsfont}
\usepackage{epsfig}
\usepackage{epstopdf}
\usepackage{physics}
\usepackage{microtype}

\usepackage{graphicx}

\usepackage{hyphenat}
\usepackage{hyperref}
\usepackage{placeins}	
\usepackage{tikz}
\usepackage{xcolor}
\usepackage{color}
\usepackage{xspace}
\usepackage{accents}

\usepackage{array}
\newcolumntype{C}[1]{>{\centering\arraybackslash}m{#1}}
\usepackage{overpic}

\newcommand\Hc{\mbox{H.\,c.}\xspace}
\newcommand\etal{\mbox{\emph{et\penalty50\ al.}}\xspace}

\renewcommand{\vec}[1]{\boldsymbol{\mathbf{#1}}}
\renewcommand{\vec}[1]{\boldsymbol{\mathbf{#1}}}

\usepackage[normalem]{ulem}

\usepackage{booktabs}
\usepackage{diagbox}
\usepackage{siunitx}
\AtBeginDocument{
	\heavyrulewidth=.08em
	\lightrulewidth=.05em
	\cmidrulewidth=.03em
	\belowrulesep=.65ex
	\belowbottomsep=0pt
	\aboverulesep=.4ex
	\abovetopsep=0pt
	\cmidrulesep=\doublerulesep
	\cmidrulekern=.5em
	\defaultaddspace=.5em
}
\def\dimerlefttriangle{\tikz[baseline=.1ex,scale=1.5]{
        \fill (0,1ex) circle (.5pt) coordinate (A);
        \fill (0,0) circle (.5pt) coordinate (B);
        \fill (.5ex,-.3ex) circle (.5pt) coordinate (C);
        \fill (1.366ex,.2ex) circle (.5pt) coordinate (D);
        \fill (.5ex,1.3ex) circle (.5pt) coordinate (E);
        \fill (1.366ex,.8ex) circle (.5pt) coordinate (F);
        \draw [line width=.9pt] (A) -- (B);
        \draw [line width=.9pt] (C) -- (D);
        \draw [line width=.9pt] (E) -- (F);}
}

\def\dimerrighttriangle{\tikz[baseline=.1ex,xscale=-1.5,yscale=1.5]{
	\fill (0,1ex) circle (.5pt) coordinate (A);
	\fill (0,0) circle (.5pt) coordinate (B);
	\fill (.5ex,-.3ex) circle (.5pt) coordinate (C);
	\fill (1.366ex,.2ex) circle (.5pt) coordinate (D);
	\fill (.5ex,1.3ex) circle (.5pt) coordinate (E);
	\fill (1.366ex,.8ex) circle (.5pt) coordinate (F);
	\draw [line width=.9pt] (A) -- (B);
	\draw [line width=.9pt] (C) -- (D);
	\draw [line width=.9pt] (E) -- (F);}
}

\begin{document}

\title{Quantum criticality of the transverse-field Ising model with long-range interactions on triangular-lattice cylinders}
\author{Jan Koziol}
\affiliation{Lehrstuhl f\"ur Theoretische Physik I, Staudtstra{\ss}e 7, Universit\"at Erlangen-N\"urnberg, D-91058 Erlangen, Germany}
\author{Sebastian Fey}
\affiliation{Lehrstuhl f\"ur Theoretische Physik I, Staudtstra{\ss}e 7, Universit\"at Erlangen-N\"urnberg, D-91058 Erlangen, Germany}
\author{Sebastian C. Kapfer}
\affiliation{Lehrstuhl f\"ur Theoretische Physik I, Staudtstra{\ss}e 7, Universit\"at Erlangen-N\"urnberg, D-91058 Erlangen, Germany}
\author{Kai Phillip Schmidt}
\affiliation{Lehrstuhl f\"ur Theoretische Physik I, Staudtstra{\ss}e 7, Universit\"at Erlangen-N\"urnberg, D-91058 Erlangen, Germany}

\begin{abstract}
To gain a better understanding of the interplay between frustrated
long-range interactions and zero-temperature quantum fluctuations,
we investigate the ground-state phase diagram of the transverse-field
Ising model with algebraically-decaying long-range Ising interactions
on quasi one-dimensional infinite-cylinder triangular lattices. Technically,
we apply various approaches including low- and high-field
series expansions. For the classical long-range Ising model, we investigate
cylinders with an arbitrary even circumference. We show the occurrence of
gapped stripe-ordered phases emerging out of the infinitely-degenerate
nearest-neighbor Ising ground-state space on the two-dimensional triangular lattice.
Further, while cylinders with circumferences $6$, $10$, $14$ et cetera are always in
the same stripe phase for any decay exponent of the long-range Ising interaction, the family
of cylinders with circumferences $4$, $8$, $12$ et cetera displays a phase transition
between two different types of stripe structures. For the full long-range
transverse-field Ising model, we concentrate on cylinders with circumference four and six.
The ground-state phase diagram consists of several quantum phases in both cases
including an $x$-polarized phase, stripe-ordered phases, and clock-ordered phases which
emerge from an order-by-disorder scenario already present in the nearest-neighbor model.
In addition, the generic presence of a potential intermediate gapless phase with
algebraic correlations and associated Kosterlitz-Thouless transitions is discussed for both cylinders. 
\end{abstract}

\maketitle

%Introduction
%%%%%%%%%%%%%%%%%%%%%%%%%%%%%%%%%%%%%%%%%%%%%%%%%%%%%%%%%%%%%%%%%%%%%%%%%%%%%%%%%%%%%%%%%%%%
The search for exotic phases of quantum matter and the identification of unconventional quantum-critical behavior is one prominent theme in current research on correlated quantum many-body systems. One important knob to trigger such exotic quantumness is frustration, which can either be present due to the lattice geometry like in antiferromagnetic quantum magnets on the triangular, Kagome or pyrochlore lattice containing odd loops or result from conflicting interactions like, most prominently, in Kitaev's honeycomb model \cite{Kitaev2006} realizing a topologically-ordered quantum spin liquid. Typically, all the paradigmatic models studied in this context have short-range interactions. 

There are, however, also important instances where long-range interactions give rise to non-trivial properties, e.g.~in the spin-ice systems where the occurrence of magnetic monopoles is a consequence of the long-range dipole-dipole interaction \cite{Castelnovo2008}, or in ferromagnetic, unfrustrated long-range transverse-field Ising models (LRTFIMs) where critical exponents can vary continuously as a function of the strength of an algebraically decaying Ising interaction \cite{Nagle1970,Fisher1972,Dutta2001,Knap2013,Fey2016,Defenu2017,Fey2019}. It is therefore natural to investigate the interplay between frustration and long-range interactions, which we expect to result in unconventional quantum behavior. Further, this interplay is of direct relevance for experimental systems, most importantly in quantum simulators with Rydberg atoms displaying an effective van-der-Waals coupling \cite{Schauss2012} as well as with trapped cold ions allowing to realize a LRTFIM with tunable interactions on the geometrically frustrated triangular lattice \cite{Britton2012, Islam2013, Bohnet2016}.

The transverse-field Ising model with algebraically decaying long-range interaction on the triangular lattice represents therefore the paradigmatic model to study the interplay of frustration and long-range interactions. While several studies have focused on the same model on a one-dimensional chain \cite{Koffel2012,Knap2013,Fey2016,Sun2017,Horita2017,Vanderstraeten2018} including the frustrated antiferromagnetic case, there are less works on the two-dimensional problem on the triangular lattice \cite{Humeniuk2016,Fey2019}, which also reflects the higher complexity from a numerical perspective. Using the recently developed high-order series expansion approach for such systems \cite{Fey2019}, it was found that the system displays the same quantum phase transition as the short-range nearest-neighbor model as long as the decay exponent of the long-range Ising interactions is not too small: There is a 3D-XY transition separating the high-field polarized phase from a clock-ordered phase \cite{Moessner2001,Moessner2003,Powalski2013}, which results from an order-by-disorder scenario about the Ising limit possessing an extensive ground-state degeneracy of spin-ice states \cite{Moessner2001}. However, the situation for more slowly decaying Ising interactions is far less understood. This lead Saadatmand \etal\cite{Saadatmand2018} to investigate the LRTFIM on a quasi-one-dimensional triangular cylinder lattice with circumference six by infinite-size density matrix renormalization group (iDMRG) calculations. Interestingly, apart from a similar type of clock order as well as a trivial polarized phase, a symmetry-broken columnar stripe phase is present in the ground-state phase diagram. However, the limit of a small magnetic field has not been studied in great detail.    

The latter findings have motivated the current paper where we investigate the ground-state phase diagram of the LRTFIM in a comprehensive fashion. To this end, we first concentrate on the classical long-range Ising model (LRIM) for a generic circumference of even length. We find that cylinders with circumference $4$, $8$, $12$, et cetera display two distinct striped phases as a function of the long-range interaction, while cylinders with circumference $6$, $10$, $14$, and so on always realize the same stripe structure for all long-range Ising interactions, which is, however, distinct from the one found numerically in Ref.~\onlinecite{Saadatmand2018}. Next we focus on the cylinder with minimal circumference of both families, namely of length $4$ and $6$, and study the ground-state phase diagram of the full LRTFIM.

The paper is structured as follows: We start by introducing the model and discussing several limiting cases as well as different representations in Sect.~\ref{sec:model}. In Sect.~\ref{sec:methods} an overview of the implemented methods used to derive the ground-state phase diagrams of the LRTFIM on the YC($4$) and YC($6$) is given. Sect.~\ref{sec:pure_long_range_ising_model} contains results for the LRIM without a magnetic field. The ground-state phase diagram for the full LRTFIM is presented and discussed for both lattices in Sect.~\ref{sec:phase_diagrams}. Finally, we conclude our work in Sect.~\ref{sec:conclusion} which includes a discussion on the presence of critical intermediate phases in the ground-state phase diagram of the LRTFIM.

%Model
%%%%%%%%%%%%%%%%%%%%%%%%%%%%%%%%%%%%%%%%%%%%%%%%%%%%%%%%%%%%%%%%%%%%%%%%%%%%%%%%%%%%%%%%%%%%
\section{Model} 
\label{sec:model}

The Hamiltonian of the LRTFIM is given by
\begin{align}
  \mathcal{H} = \frac{J}{2}\sum_{{\bf i} \neq {\bf j}} \frac{1}{|{\bf i}-{\bf j}|^\alpha}\sigma_{{\bf i}}^z\;\sigma_{{\bf j}}^z -h\sum_{{\bf j}}\sigma_{{\bf j}}^x  ~ , \label{eq:H_tfim_orig}
\end{align}
with Pauli matrices \(\sigma_{{\bf i}}^{x/z}\) describing spins-1/2 located on lattice sites \({\bf i}\), the transverse field \(h>0\), and the antiferromagnetic coupling constant \(J>0\). Tuning the positive parameter \(\alpha\) changes the long-range behavior of the Ising interaction from an all-to-all coupling $\alpha=0$ up to the nearest-neighbor case $\alpha=\infty$.

\begin{figure}
    \centering
    \includegraphics{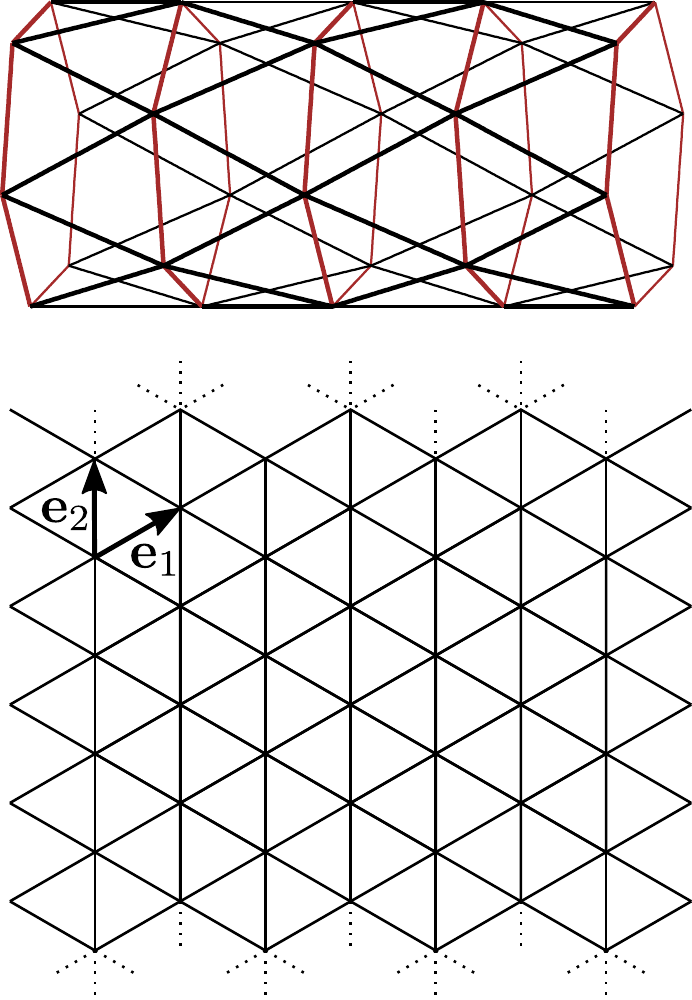}
    \caption{Illustration of the YC6 lattice. Upper panel: YC$(6)$ embedded in 3D to illustrate the periodic boundary conditions. Lower panel: Illustration as a cutout of the 2D triangular lattice with periodic boundary conditions denoted with dashed lines. The vectors $\vec{e}_1$ and $\vec{e}_2$ are the unit vectors which span the whole triangular cylinder.}
    \label{fig:cylinder_2d_and_3d_illustration}
\end{figure}

The triangular cylinders consist of rings with circumference $n$ which are coupled in the direction of infinite extension to form a triangular lattice as illustrated in Fig.~\ref{fig:cylinder_2d_and_3d_illustration}. According to the number of spins per ring $n$, these lattices are labeled YC($n$), while we focus on $n\in\{4,6\}$ for most parts in this paper. In Fourier space the momentum orthogonal to the infinite cylinder extension becomes discrete while it is continuous in the other direction due to the infinite extension.

%Figure 1: Illustration YC6
%%%%%%%%%%%%%%%%%%%%%%%%%%%%%%%%%%%%%%%%%%%%%%%%%%%%%%%%%%%%%%%%%%%%%%%%%%%%%%%%%%%%%%%%%%%%
%%%%%%%%%%%%%%%%%%%%%%%%%%%%%%%%%%%%%%%%%%%%%%%%%%%%%%%%%%%%%%%%%%%%%%%%%%%%%%%%%%%%%%%%%%%%

It is convenient to rewrite the summations in Eq.~\eqref{eq:H_tfim_orig} with respect to the characteristics of the YC$(n)$ cylinders. A lattice site can either be addressed by unit vectors $\vec{e}_1$ and $\vec{e}_2$ (see Fig.~\ref{fig:cylinder_2d_and_3d_illustration}) like for the 2D triangular lattice with periodic boundary conditions, or by the ring $R\in\mathbb{Z}$ (highlighted in red in Fig.~\ref{fig:cylinder_2d_and_3d_illustration}) the site is located on and the position within the ring $\nu\in\{ 0,1,2,...,n-1 \}$.
The Hamiltonian Eq.~\eqref{eq:H_tfim_orig} can then be expressed as 
\begin{equation}
	\mathcal{H}=\frac{J}{2}\sum_{R,R'}\sum_{\nu,\nu'=0}^{n-1}d_{R,\nu}^{R',\nu'}(\alpha) \,\sigma^z_{R,\nu}\sigma^z_{R',\nu'}-h\sum_{R}\sum_{\nu=0}^{n-1}\sigma^x_{R,\nu}
\end{equation}
with $d_{R,\nu}^{R',\nu'}(\alpha)$ representing the $\alpha$-dependent long-range Ising interaction between the sites $(R,\nu)$ and $(R',\nu')$. Obviously one has $d_{R,\nu}^{R,\nu}(\alpha)=0$, since no self-interactions are present. Note that we measure distances in the planar geometry as illustrated in the lower panel of Fig.~\ref{fig:cylinder_2d_and_3d_illustration}.

\begin{figure}[tb]
	\centering
	\includegraphics[width=.9\columnwidth]{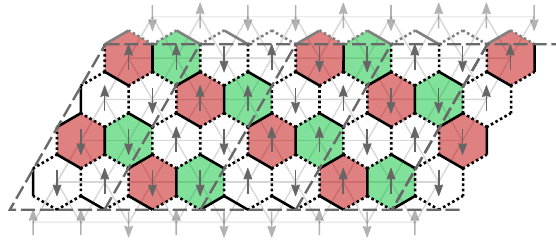}
	\caption{Illustration of the maximally-flippable state of YC$(4)$ in the quantum-dimer model. Flippable plaquettes are shown as red and green hexagons. The original triangular cylinder is shown in the background together with a spin configuration resulting in the displayed dimer configuration. The periodic boundary is reflected in lighter gray.}
	\label{fig:dimer_covering_YC4}
\end{figure}

\begin{figure}[tb]
	\centering
	\includegraphics[width=\columnwidth]{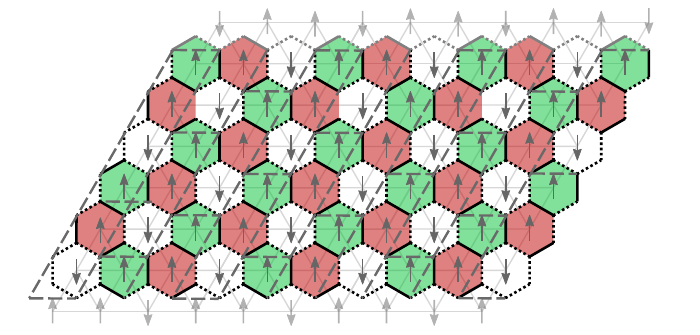}
	\caption{Illustration of the maximally-flippable state of YC$(6)$ in the quantum-dimer model. Flippable plaquettes are shown as red and green hexagons. The original triangular cylinder is shown in the background together with a spin configuration resulting in the displayed dimer configuration. The periodic boundary is reflected in lighter gray.}
	\label{fig:dimer_covering_YC6}
\end{figure}

The LRTFIM has several interesting limiting cases. Studying them can lead to a better intuition for the involved physics and they are later used as starting points for perturbative calculations. For $h=\infty$ the system is in an $x$-polarized phase where all spins align in direction of the transverse field independently of the lattice. This is a commonly-used starting point for high-field perturbative approaches due to the clear reference state including the long-range case $\alpha<\infty$ \cite{Fey2016,Fey2019}. For $\alpha=\infty$ and $n=\infty$ the system corresponds to the nearest-neighbor transverse-field Ising model \mbox{(NNTFIM)} with an infinitely-degenerate ground-state space for $h=0$ on the 2D triangular lattice. Here every state is a ground state of the system that obeys the rule that on every triangle there are one ferro- and two antiferromagnetic bonds. For the 2D triangular lattice it is known that this highly-degenerate nearest-neighbor ground-state space is not stable against an infinitesimal transverse field $h$ which leads to an order-by-disorder scenario inducing a clock-ordered phase. This is most easily seen by performing first-order degenerate perturbation theory in $h/J$ yielding an effective quantum dimer model on the dual honeycomb lattice \cite{Moessner2001}
\begin{equation}\label{eq:qdm}
	\mathcal{H}_{\text{QDM}} = -h\sum_{\vec{\nu}}\left(\ket{\dimerlefttriangle}_{\vec{\nu}}\bra{\dimerrighttriangle}_{\vec{\nu}} +\Hc\right)
\end{equation}
up to an irrelevant constant. The sum over $\vec{\nu}$ runs over all hexagonal plaquettes of the dual honeycomb lattice as illustrated in Figs.~\ref{fig:dimer_covering_YC4} and \ref{fig:dimer_covering_YC6} for the YC($4$) and YC($6$) lattice. In this representation bonds with ferromagnetically-oriented neighbor spins are interpreted as a dimer in contrast to bonds with antiferromagnetically-oriented spins. In general, when acting on a spin with the field term, all dimers of a hexagonal plaquette are flipped to non-dimers and vice versa. If the total number of dimers on a plaquette remains unchanged, a different ground state is produced. Therefore, the effective \mbox{Hamiltonian \eqref{eq:qdm}} describes quantum fluctuations in the ground-state space between the two configurations on plaquettes with three ferro- and antiferromagnetic bonds. Plaquettes having such a configuration are called \emph{flippable}. Obviously, states with a maximum number of flippable plaquettes are selected energetically. For the 2D triangular lattice the resulting ground state is the above mentioned clock order which breaks the translational symmetry of the lattice. The same mechanism takes place on the cylindric triangular lattices. For the nearest-neighbor model we find a quasi momentum of the clock-ordered state of $(5\pi/4,\pi/2)$ [$(2\pi/3,-2\pi/3)$] for the YC($4$) [YC($6$)] lattice (see Figs.~\ref{fig:dimer_covering_YC4} and \ref{fig:dimer_covering_YC6}). Note that for the YC($6$) cylinder this corresponds to the same order as in 2D while the clock order for YC($4$) is distinct, since the unit cell of the 2D clock order does not fit on this cylinder. For the NNTFIM on the 2D triangular lattice, the quantum phase transition between the clock order and the high-field $x$-polarized phase is known to be second order in the 3D-XY universality class \cite{Moessner2001,Moessner2003,Powalski2013}. For finite $\alpha<\infty$, the nature of this quantum phase transition is unchanged for all $\alpha\gtrapprox 2.5$ \cite{Fey2019,Humeniuk2016}.

Saadatmand \etal investigated the LRTFIM on the YC$(6)$ cylinder for $\alpha\in(1,5)$ as well as the NNTFIM using iDMRG \cite{Saadatmand2018}. They find for $\alpha>2.40(5)$ the same quantum phases as for the 2D triangular lattice, with a transition between the clock order and the $x$-polarized phase. The critical point for the NNTFIM is located at $h_c=1.5(1)\,J$ \cite{Saadatmand2018}. For $\alpha<2.40(5)$, they observe a direct phase transition from the $x$-polarized phase into a different ordered phase, which we will call zigzag-stripe phase (see Fig.~\ref{fig:stripes} for an illustration). Interestingly, our investigation of the pure LRIM for $h=0$ confirms the appearance of stripe-ordered phases (for any finite $\alpha$), although we find a different stripe order to be realized which we attribute to the chosen unit cell in Ref.~\onlinecite{McCulloch_Private}.

Finally, the LRTFIM Eq.~\eqref{eq:H_tfim_orig} reduces to a fully connected graph with equal all-to-all coupling in the limit $\alpha=0$. It is then convenient to rewrite the Hamiltonian by introducing the total spin $\sigma_{\text{tot}}^{x/z}=\sum_{{\bf i}}\sigma_{{\bf i}}^{x/z}$ and the total number of spins $N$ with $N\rightarrow \infty$ to obtain
\begin{align}
	\mathcal{H} = -h\sigma^x_{\text{tot}}+\frac{J}{2}(\sigma^z_{\text{tot}})^2-\frac{J}{2}N\quad ,
  	\label{eq:H_tfim_meanfield}
\end{align}
which immediately shows that every state with vanishing magnetization is a ground state for $h=0$ so that a large degeneracy results. For all finite transverse fields $h>0$ the system breaks this degeneracy and is directly located in the $x$-polarized phase \cite{Humeniuk2016}. One then regains the full LRTFIM by adding
	\begin{equation}
	\frac{J}{2}\sum_{{\bf i} \neq {\bf j}} \left(\frac{1}{|{\bf i}-{\bf j}|^\alpha}-1\right)\,\sigma_{{\bf i}}^z\;\sigma_{{\bf j}}^z\ . 
		\label{eq:smallalphaunperturbed}
	\end{equation}
        to the all-to-all limit Eq.~\eqref{eq:H_tfim_meanfield}. It is therefore also possible to consider the infinitely degenerate limit $h=0$ and $\alpha=0$ as perturbative starting point. For the long-range Ising interactions this demands a Taylor expansion of Eq.~\eqref{eq:smallalphaunperturbed} giving in leading order in $\alpha$
	\begin{equation}
		 -\alpha\frac{J}{2}\sum_{{\bf i} \neq {\bf j}}\log(|{\bf i}-{\bf j}|)\,\sigma_{{\bf i}}^z\;\sigma_{{\bf j}}^z \ .
		\label{eq:smallalphalogperturb}
	\end{equation}
This perturbation describes an extensive ferromagnetic LRIM with logarithmically increasing Ising interaction strength.

Our goal is to determine the full ground-state phase diagram of the LRTFIM on the YC$(4)$ and YC$(6)$ cylinder. To this end, we apply several perturbative expansions in the $x$-polarized phase and in the stripe- and clock-ordered phases. The technical aspects are discussed next.

%Figures 2-4: Illustration QDM/clock order YC4/YC6
%%%%%%%%%%%%%%%%%%%%%%%%%%%%%%%%%%%%%%%%%%%%%%%%%%%%%%%%%%%%%%%%%%%%%%%%%%%%%%%%%%%%%%%%%%%%

%%%%%%%%%%%%%%%%%%%%%%%%%%%%%%%%%%%%%%%%%%%%%%%%%%%%%%%%%%%%%%%%%%%%%%%%%%%%%%%%%%%%%%%%%%%%

%Methods
%%%%%%%%%%%%%%%%%%%%%%%%%%%%%%%%%%%%%%%%%%%%%%%%%%%%%%%%%%%%%%%%%%%%%%%%%%%%%%%%%%%%%%%%%%%%
\section{Methods}\label{sec:methods}
To map out the ground-state phase diagram of the LRTFIM on the triangular cylinders we set up several methods for the calculation of the ground-state energy as well as the elementary excitation energy in the different quantum phases just introduced in the last section. First, we describe the perturbative expansion of both quantities about the high-field limit in the $x$-polarized phase and the Pad\'{e} extrapolation of the corresponding series. Then, we explain the perturbative low-field expansion about the stripe-ordered ground states of the LRIM in $h/J$. Finally, the perturbative evaluation of the clock-ordered ground state in the presence of the transverse field and the long-range Ising interaction is explained.

\subsection{High-field expansion}\label{ssec:high_field_expansion}
%%%%%%%%%%%%%%%%%%%%%%%%%%%%%%%%%%%%%%%%%%%%%%%%%%%%%%%%%%%%%%%%%%%%%
The high-field expansion for the LRTFIM is most efficiently done by combining a white-graph expansion \cite{Coester2015} for a perturbative continuous unitary transformation (PCUT) \cite{Knetter2000,Knetter2003} with Markov-chain Monte Carlo (MCMC) as shown for two-dimensional LRTFIMs recently in Ref.~\onlinecite{Fey2019}. For details of this approach we therefore refer to Refs.~\onlinecite{Fey2016,Fey2019} where this approach was discussed before. Here, we only concentrate on the essential aspects.

Using the Matsubara-Matsuda transformation \mbox{\(\sigma_{\vec j}^z = \hat b_{\vec j}^\dagger+\hat b_{\vec j}^{\phantom{\dagger}}\)} and \mbox{\(\sigma_{\vec j}^x=1-2\hat{n}_{\vec j}\)} \cite{Matsubara1956}, the LRTFIM Eq.~\eqref{eq:H_tfim_orig} can be written, up to the constant $-N/2$, in a quasi-particle (QP) language as
\begin{align}
	  \frac{\mathcal{H}}{2h} &=
	  	\sum_{R}\sum_{\nu=0}^{n-1} \hat{n}_{R,\nu}
	  	- \frac{\lambda}{2}\sum_{R,R'}\sum_{\nu,\nu'=0}^{n-1} d_{R,\nu}^{R',\nu'}(\alpha)  \notag\\
	  		&\quad\quad\quad\left( \hat b^\dagger_{R,\nu} \hat b^\dagger_{R',\nu'} 
	  			+ \hat b^\dagger_{R,\nu} \hat b^{\phantom{\dagger}}_{R',\nu'} 
	  			+ {\rm H.c.}\right)~, \label{eq:H_tfim_orig_boson}
\end{align}
% \begin{equation}
% 	  \frac{\mathcal{H}}{2h} =\sum_{\vec{i}} \hat{n}_{\vec i} - \frac{\lambda}{2}\sum_{\vec i\neq\vec j} \frac1{|\vec i-\vec j|^\alpha}\left( \hat b^\dagger_{\vec i} \hat b^\dagger_{\vec j} + \hat b^\dagger_{\vec i} \hat b^{\phantom{\dagger}}_{\vec j} + {\rm H.c.}\right) \, , \label{eq:H_tfim_orig_boson}
% \end{equation}
where $\hat{n}_{R,\nu}=\hat{b}^\dagger_{R,\nu}\hat{b}^{\phantom{\dagger}}_{R,\nu}$ counts the number of hardcore bosons on site $(R,\nu)$ and \(\lambda\equiv J/(2h)\) is the expansion parameter. The PCUT using a white-graph expansion scheme \cite{Coester2015} maps this Hamiltonian, order by order in $\lambda$, to an effective block-diagonal Hamiltonian \(\mathcal{H}_{\text{eff}}\), which preserves the total number of QP's, i.e.~, $[\mathcal{H}_{\text{eff}},\mathcal{Q}]=0$ with \(\mathcal{Q}\equiv \sum_{R,\nu}\hat{n}_{R,\nu}\).

Here we focus on the zero- and one-QP block of $\mathcal{H}_{\text{eff}}$ so that we have access to the ground-state energy per site $e_0^{\mathrm{pol}}\equiv E_0^{\mathrm{pol}}/N$ as well as to the one-QP excitation energies. The effective Hamiltonian in the one-QP sector reads
\begin{equation}
	\mathcal{H}_{\rm eff}^{\rm 1QP} = E_0^{\mathrm{pol}} + \sum_{R,R'}\sum_{\nu,\nu'=0}^{n-1} a_{\vec{\delta}} \left( \hat{b}^\dagger_{R,\nu} \hat b^{\phantom{\dagger}}_{R',\nu'}+{\rm H.c.}\right) ~. \label{eq:H_eff_1QP}
\end{equation}
% \begin{equation}
% 	\mathcal{H}_{\rm eff}^{\rm 1QP} = E_0 + \sum_{\vec i,\vec{\delta}} a_{\vec\delta} \left( \hat{b}^\dagger_{\vec i} \hat b^{\phantom{\dagger}}_{\vec i + \vec\delta}+{\rm H.c.}\right) ~. \label{eq:H_eff_1QP}
% \end{equation}
with the ground-state energy $E_0^{\mathrm{pol}}$ and the one-QP hopping amplitudes $a_{\vec\delta}\equiv a_{R,\nu}^{R',\nu'}$, where $\vec{\delta}$ denotes the vector between lattice sites $(R,\nu)$ and $(R',\nu')$. 

Exploiting the translational invariance along the infinite cylinder direction, a Fourier transformation with respect to the unit vector $\vec{e}_1$ depicted in Fig.~\ref{fig:cylinder_2d_and_3d_illustration} transforms Eq.~\eqref{eq:H_eff_1QP} to
\begin{equation}
	\mathcal{H}_{\rm eff}^{\rm 1QP} = E^{\rm pol}_0+\sum_{k_1}\sum_{\nu,\nu'=0}^{n-1} \omega_{\nu,\nu'}(k_1)\,\hat{b}^\dagger_{k_1,\nu} \hat b^{\phantom{\dagger}}_{k_1,\nu'}~,
\end{equation}
where $k_1\in\mathds{R}$ is the one-dimensional quasi-momentum in the cylinder direction. The matrix elements $\omega_{\nu,\nu'}(k_1)$ contain information of the QP hopping from site $\nu$ to site $\nu'$ of the $n$-site ring. The one-QP Hamiltonian $\mathcal{H}_{\rm eff}^{\rm 1QP}$ can be fully diagonalized with the help of the discrete translational invariance in the finite periodic cylinder direction in $\vec{e}_2$-direction so that
\begin{equation}
	\mathcal{H}_{\rm eff}^{\rm 1QP} = E^{\rm pol}_0+\sum_{\vec{k}}\omega (\vec{k})\,\hat{b}^\dagger_{\vec{k}} \hat b^{\phantom{\dagger}}_{\vec{k}}
\end{equation}
defining the two-dimensional quasi-momentum \mbox{$\vec{k}=(k_1,k_2)$} with $k_2\in\{0,\ldots,2\pi/n\}$.

The one-QP hopping amplitudes \(a_{\vec\delta}\) are calculated on minimal graphs in the white-graph expansion and analytically exact for any lattice. To get the perturbative series in the bulk limit the amplitudes need to be embedded into the lattice. The embedding consists of a summation of the contributions of the hopping elements from all possible configurations of each graph on the lattice in a given order. As a consequence, each matrix element $\omega_{\nu,\nu'}(k_1)$ as well as each one-QP energy $\omega (\vec{k})$ is given as a high-dimensional nested sum. These summations are most efficiently evaluated numerically by a Markov-chain Monte Carlo (MCMC) method as we demonstrated recently \cite{Fey2019}. We indeed implemented two schemes. First, we sampled each element of $\omega_{\nu,\nu'}(k_1)$ separately and then calculate the one-particle dispersion $\omega (\vec{k})$ in a subsequent step. To this end, we extended the code from Ref.~\onlinecite{Fey2019} to arbitrary unit cells. Second, we directly sampled with MCMC the one-particle dispersion $\omega (\vec{k})$ for a fixed $\vec{k}$. Numerically, it turned out that the second approach yields smaller error bars and we therefore concentrate on this approach in the following. In both cases, for each perturbative order $r$, a separate MCMC calculation for all graphs with $\mu\in[2,r+1]$ vertices is done. For each MCMC calculation up to 80 runs with different random number generator seeds are computed to obtain an error estimation from the standard deviation of the mean sum value.

Using this approach, we calculated PCUT results for the ground-state energy per site $e_0^{\rm pol}$ and the one-QP dispersion $\omega (\vec{k})$ up to order 10 in the thermodynamic limit with high accuracy. Even for the demanding limit of small values of $\alpha$ we get standard deviations in the highest-order coefficientes of a low single-digit percentage magnitude. In the one-QP sector, we are mostly interested in the one-QP gap $\Delta\equiv {\rm min}_{\vec{k}}\,\omega (\vec{k})$. In the numerical evaluation we therefore focus on the momenta $(2\pi/3,-2\pi/3)$ [$(5\pi/4,\pi/2)$] for the YC($6$) [YC($4$)] lattice, where the gap between the $x$-polarized and the clock-ordered phase closes at the phase transition in the nearest-neighbor limit. Further, we consider the excitation energies at the momenta of the relevant stripe-ordered phases which are introduced in Subsect.~\ref{ssec:low_field_expansion}. As the value of the parameter $\alpha$ needs to be fixed for the numerical MCMC evaluation, we compute perturbative series for a set of fixed values of $\alpha$ in the range \([1.5,10]\). 

In the end, the resulting series for the one-QP gap are extrapolated using Pad{\'e} extrapolations \cite{Guttmann1989} in order to enlarge the convergence radius. The extrapolants are then used to obtain the critical values $\lambda_{\text{c}}$ of the phase transition for fixed $\alpha$ where the one-QP gap closes.

\subsection{Pad\'{e} extrapolations}\label{ssec:pade}

Pad\'{e} and Dlog-Pad\'{e} approximations are standard methods in the field of series expansions as they allow for an evaluation of the series beyond their original radius of convergence \cite{Guttmann1989}. Therefore, they are well-suited to study our results from the high-field expansion. While Pad\'{e} expansions are generally used for extrapolating ground-state energies \cite{Roechner2016}, the criticality and location in parameter space of a second-order phase transition is usually studied using Dlog-Pad\'{e} extrapolations of the one-QP energy gap, because it is capable of incorporating an algebraic gap-closing. As discussed below, we do not expect a second-order phase transition and consequently estimate the phase-transition point via Pad\'{e} extrapolations. 

The perturbation series 
\begin{equation}
	F(\lambda)=\sum_{m\geq 0}^{r} c_m \lambda^m=c_0+c_1\lambda+c_2\lambda^2+\dots c_{r}\lambda^r,
\end{equation}
with $\lambda\in \mathbb{R}$ and $c_m \in \mathds{R}$ is interpreted as a Taylor expansion of a rational function
\begin{equation}
	G^{L/M}(\lambda) = \frac{p_0 + p_1 \lambda + p_2 \lambda^2 +\cdots+p_L \lambda^L}{1 + q_1 \lambda + q_2 \lambda^2 +\cdots+q_M \lambda^M}~.
\end{equation}
Comparing the series expansion of $G^{L/M}(\lambda)$ with the original series $F(\lambda)$ one obtains a linear system of equations that can be solved for a given parameter set $(L,M)$. The parameters $L,M\in\mathds{N}$ fulfill the condition $L+M=r$ for an extrapolation in order $r$. Typically the diagonal extrapolations where $|L-M|$ is small give the best results.  Extrapolations with unphysical singularities need to be sorted out, as well as defective Pad\'{e} extrapolants that have a singularity at the same point in the numerator and denominator which effectively cancel out. The phase transition point $\lambda_\mathrm{c}$ studied in the present paper is found by calculating the zeros of the Pad\'{e} extrapolation of the 1QP gap. To locate the phase transition, we calculate several values for the critical parameter $\lambda_\mathrm{c}$ for different selected combinations $(L,M)$ out of \mbox{$\{(3,5),(4,4),(5,3),(4,5),(5,4),(4,6),(5,5),(6,4)\}$} for non-defective Pad\'{e} extrapolants. In the phase diagrams presented in Sect.~\ref{sec:phase_diagrams} we show the mean and standard deviation of the extrapolations. If the standard deviation is zero only a single extrapolation could be selected.

\subsection{Low-field expansion}\label{ssec:low_field_expansion}
%%%%%%%%%%%%%%%%%%%%%%%%%%%%%%%%%%%%%%%%%%%%%%%%%%%%%%%%%%%%%%%%%%%%%
The opposite limit of small transverse fields $h/J\ll 1$ can be also be treated by high-order series expansions for $\alpha<\infty$, since the extensive ground-state degeneracy for the nearest-neighbor case $\alpha=\infty$ is lifted by the long-range Ising interaction. We therefore have determined the ground state of the LRIM as a function of $\alpha$ by considering large but finite triangular cylinders YC($n$), with periodic boundary conditions, for general even $n$. These findings are outlined and discussed in Sect.~\ref{sec:pure_long_range_ising_model}. As a result of these calculations we find that the pure LRIM realizes different types of ordered stripe structures depending on $n$ and $\alpha$ and we can determine the associated ground-state energy per site $e_0^{\rm stripe}$ in units of~$J$ by considering finite cylinders of the order of $N=10^5$ spins.

\begin{figure}
	\centering
	\includegraphics{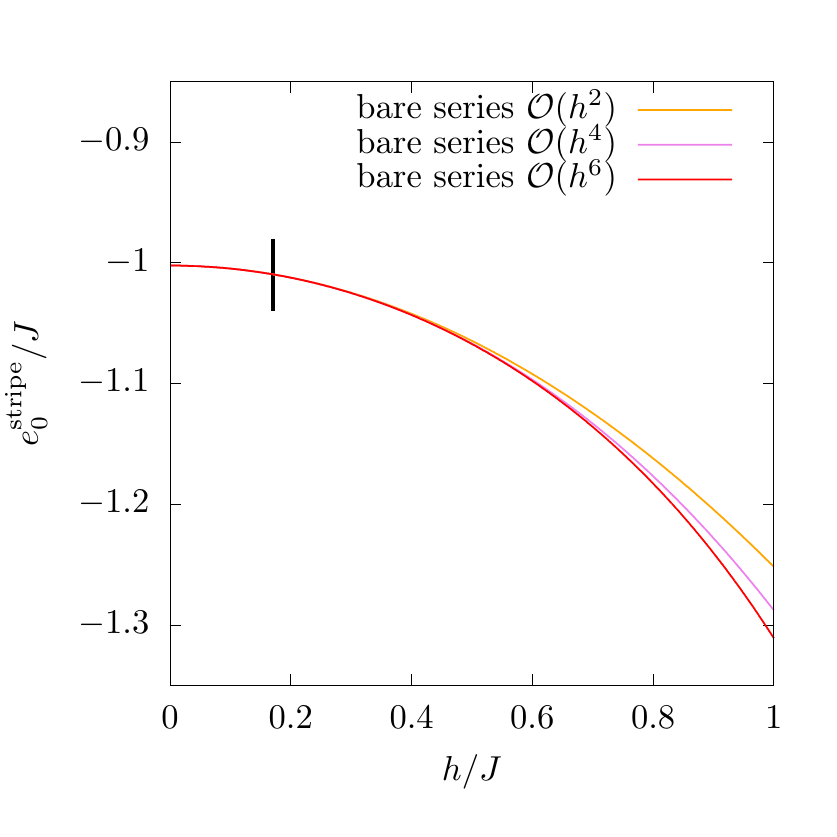}
	\caption{Bare series of the ground-state energy per site $e_0^{\rm stripe}$ for orthogonal stripes as a function of $h/J$ for $\alpha=6$ for the YC$(4)$ cylinder. The bare series is converged up to $h/J\approx 0.6$. The behavior of other stripe configurations as well as other $\alpha$-values is similar. The calculated phase transition into the clock order, visualized by the black vertical line, takes place at approximately $0.17\,h/J$.}
	\label{fig:convergence_low_field}
\end{figure}

These ordered stripe structures represent gapped phases which allows us to set up a high-order (non-degenerate) series expansion about the zero-field ground state. To this end, we apply Takahashi's perturbation theory \cite{Takahashi1997} in real space and we obtain the ground-state energy per site $e_0^{\rm stripe}$ for various stripe structures up to order six in the parameter $h/J$. To do this, we calculate the even-order contributions directly by evaluating the expectation value of the perturbation-operator sequences with respect to the considered classical stripe state. We stress that only even orders are present in the low-field expansion of $e_0^{\rm stripe}$, while odd orders vanish exactly. This originates from a double-touch property, because each excitation created locally by the perturbing magnetic field in virtual states has to be destroyed by acting again with the magnetic field on the same site. As a consequence, every site has to be touched an even number of times by the magnetic field to get a non-vanishing result. The perturbation operator sequences in $r$-th order read as follows \cite{Takahashi1997}
\begin{equation}
	\hat{P}\mathcal{V}\hat{S}^{k_1}\mathcal{V}\hat{S}^{k_2}\mathcal{V}...\mathcal{V}\hat{S}^{k_{r-1}}\mathcal{V}\hat{P}\, ,
\end{equation}
where 	$\mathcal{V}\equiv\mathcal{H}_{\rm h}$ is the perturbation, $\hat{P}$ the projection operator on the ground-state space, the resolvent $\hat{S}$ is given as
\begin{align}
 \hat{S}=\frac{(1-\hat{P})}{E_0^{\text{stripe}}-\mathcal{H}_{\rm Ising}} &\quad \mathrm{with}\quad \hat{S}^k=
	\begin{cases}
		\hat{P} & k=0\\
		\hat{S}^k & k>0 
	\end{cases} \ , 
\end{align}
and the constraint $\sum_{i=1}^{n-1}k_i=r-1$. We evaluate all contributions up to order six in $h/J$ by calculating the expectation values
\begin{align}
	\bra{{\rm stripe}}  \hat{P}\mathcal{V}\hat{S}^{k_1}\mathcal{V}\hat{S}^{k_2}\mathcal{V}\ldots\mathcal{V}\hat{S}^{k_{r-1}}\hat{P} \ket{{\rm stripe}}
\end{align}
for the classical stripe state $\ket{{\rm stripe}}$ on finite clusters with $1000n$ spins for the YC$(n)$ cylinder and by treating the perturbing magnetic field in real space. We have further reduced the summation effort by identifying the non-vanishing processes in advance and exploiting the translational invariance for the first excitation that is created by the perturbation $\mathcal{V}$. Finally, for a fixed $\alpha$ we obtain the following order-six series of the ground-state energy per site for $J=1$
\begin{align}
	e_0^{\text{stripe}}(\alpha,h) =& e_0^{\text{stripe}}(\alpha,h=0)+\rho_2^{\text{stripe}}(\alpha)\;h^2\\
				    &+\rho_4^{\text{stripe}}(\alpha)\; h^4+\rho_6^{\text{stripe}}(\alpha)\;h^6 \quad .
\end{align}
As a representative example, the bare series in order two, four, and six of the ground-state energy per site of orthogonal stripes for $\alpha=6$ are displayed in Fig.~\ref{fig:convergence_low_field}. In general, we observe that the first-order phase transition out of the stripe-ordered phase is well located in the regime where the bare series is still converged. Consequently, we do not extrapolate the series of the low-field expansions.
  
%%%%%%%%%%%%%%%%%%%%%%%%%%%%%%%%%%%%%%%%%%%%%%%%%%%%%%%%%%%%%%%%%%%%%%%%%%%%%
%%%%%%%%%%%%%%%%%%%%%%%%%%%%%%%%%%%%%%%%%%%%%%%%%%%%%%%%%%%%%%%%%%%%%%%%%%%%%

\subsection{Clock-ordered states}\label{ssec:clock_orders}
%%%%%%%%%%%%%%%%%%%%%%%%%%%%%%%%%%%%%%%%%%%%%%%%%%%%%%%%%%%%%%%%%%%%%
As already outlined above, apart from the high-field $x$-polarized phase and the low-field stripe phases for $\alpha<\infty$, one expects also clock-ordered phases in the ground-state phase diagram. These clock-ordered states are stabilized by an order-by-disorder phenomenon, i.e.~this order is selected by the quantum fluctuations induced by an infinitely small transverse field on the extensive ground-state manifold for $\alpha=\infty$ and $h=0$. We are therefore determining approximately the ground-state energy per site $e_0^{\rm clock}$ of these clock-ordered states as a function of the long-range interaction $\alpha$ and the transverse field $h$. To this end, we split the Hamiltonian 
%%%%%%%%%%%%%%%%%%%%%%%%%%%%%%%%%%%%%%%%%%%%%%%%%%%%%%%%%%%%%%%%%%%%%
\begin{equation}
  \frac{\mathcal{H}}{J}=\mathcal{H}_{\rm Ising}^{\alpha=\infty}+\frac{h}{J}\mathcal{H}_{\rm field}+\xi\Delta\mathcal{H}_{\rm Ising}\quad ,
\end{equation}
%%%%%%%%%%%%%%%%%%%%%%%%%%%%%%%%%%%%%%%%%%%%%%%%%%%%%%%%%%%%%%%%%%%%%
where the last term is defined as the difference between the full long-range Ising interaction and the nearest-neighbor contribution
%%%%%%%%%%%%%%%%%%%%%%%%%%%%%%%%%%%%%%%%%%%%%%%%%%%%%%%%%%%%%%%%%%%%%
\begin{equation}
  \Delta\mathcal{H}_{\rm Ising} \equiv \mathcal{H}_{\rm Ising} - \mathcal{H}_{\rm Ising}^{\alpha=\infty}
\end{equation}
%%%%%%%%%%%%%%%%%%%%%%%%%%%%%%%%%%%%%%%%%%%%%%%%%%%%%%%%%%%%%%%%%%%%%
and one recovers the original LRTFIM for $\xi=1$. In the following, we consider the perturbative limit \mbox{$\xi\ll h \ll J$}, i.e.~we expand about the nearest-neighbor Ising model using degenerate perturbation theory so that an effective description in terms of a quantum dimer model is appropriate. 

Indeed, due to the frustration, the ground-state subspace consists of infinitely many degenerate states which have in each elementary triangle exactly one ferromagnetic bond. Interpreting the ferromagnetic bond as the presence of a dimer on the dual honeycomb lattice, each ground state for $\alpha=\infty$ and $h=0$ can therefore be represented by a dimer covering $\ket{c}$ and the (perturbative) action of the two perturbations $\mathcal{H}_{\rm field}$ and $\Delta\mathcal{H}_{\rm Ising}$ can be captured by an effective quantum dimer model of the form
%%%%%%%%%%%%%%%%%%%%%%%%%%%%%%%%%%%%%%%%%%%%%%%%%%%%%%%%%%%%%%%%%%%%%
\begin{align}\label{eq:qdm2}
  \mathcal{H}_{\text{QDM}} = E_0& +\sum_{c}E_{c}(\xi,h)\ket{c}\bra{c}\nonumber\\
                         & -h\sum_{\vec{\nu}}\left(\ket{\dimerlefttriangle}_{\vec{\nu}}\bra{\dimerrighttriangle}_{\vec{\nu}} +\Hc\right)\,,
\end{align}
%%%%%%%%%%%%%%%%%%%%%%%%%%%%%%%%%%%%%%%%%%%%%%%%%%%%%%%%%%%%%%%%%%%%%
where the sum runs over all dimer coverings $\ket{c}$ so that $E_{c}(\xi,h)$ is the covering-dependent diagonal energy. We have determined this effective quantum dimer model up to order three in the parameters $\xi$ and $h$. The diagonal elements $E_{c}(\xi,h)$ have several contributions. In first order in $\xi$, it depends on the long-range part $\Delta\mathcal{H}_{\rm Ising}$ of the Ising interaction. Note however that this first-order contribution actually represents the exact pure \mbox{$\xi$-dependent} energy correction, since $[\mathcal{H}_{\rm Ising}^{\alpha=\infty},\Delta\mathcal{H}_{\rm Ising}]=0$. Further, $E_{c}(\xi,h)$ depends on $h^2$ and $h^2\xi$ in second- and third-order degenerate perturbation theory. The only off-diagonal term comes in first order which mediates between two different dimer coverings from the transverse field $h$ as already discussed above.

Since we consider the hierarchy \mbox{$\xi\ll h \ll J$}, it is the (dressed) maximally-flippable plaquette state (see Figs.~\ref{fig:dimer_covering_YC4} and \ref{fig:dimer_covering_YC6}), which is selected due to an infinitesimal transverse field and which gives rise to the clock-ordered phase on the YC($n$) cylinders. Calculating the ground-state energy $e_0^{\rm clock}$ as a function of $\alpha$ and $h$ is still a highly non-trivial task due to the fact that the quantum dimer model Eq.~\eqref{eq:qdm2} still lives in an infinite-dimensional Hilbert space.

We therefore have used the following scheme to calculate $e_0^{\rm clock}$ approximatively. First, we restrict the calculations to finite cylinders with the number of $8$ ($6$) rings for $n=4$ ($n=6$), which obviously yields a finite-dimensional problem. In order to focus on the most relevant dimer coverings on these finite cylinders for the description of the clock-ordered state, we generate the following reduced basis. We start from the maximally-flippable state (see Figs.~\ref{fig:dimer_covering_YC4} and \ref{fig:dimer_covering_YC6}) and then we act subsequently with the magnetic field on flippable plaquettes. This generates new states having less flippable plaquettes on which we can again act with the magnetic field to generate further dimer coverings and so on. Consequently, we iteratively construct a finite basis of dimer coverings where all states are connected to the maximally-flippable state via the magnetic field. The Hamiltonian \eqref{eq:qdm2} can then be represented as a finite matrix in this basis and the lowest eigenvalue corresponds to an approximation of $e_0^{\rm clock}$.

For the YC($4$) [YC($6$)] cylinder, the maximally-flippable state has an 8-site (3-site) unit cell. Our results are therefore computed on four and twelve unit cells for the respective cylinders. The finite cluster size has two competing effects on the calculated ground-state energy $e_0^{\rm clock}$: First, especially relevant for large $\alpha$, the limitation to a finite cylinder results in an underestimation of the quantum fluctuations introduced by the transverse field. Consequently, we expect that in the bulk-limit $e_0^{\rm clock}$  should be pushed down stronger with the magnetic field compared to our finite-size calculation. Second, the long-range Ising interaction $\Delta\mathcal{H}_{\rm Ising}$, which becomes more and more important for smaller $\alpha$, should lead to an increased energy in the bulk limit. Considering both contributions, we expect the method to underestimate the energy for small $\alpha$ and to overestimate it slightly for large $\alpha$.

% Pure Long-range Ising Model 
%%%%%%%%%%%%%%%%%%%%%%%%%%%%%%%%%%%%%%%%%%%%%%%%%%%%%%%%%%%%%%%%%%%%%%%%%%%%%%%%%%%%%%%%%%%%
\section{Pure Long-range Ising Model} % (fold)
\label{sec:pure_long_range_ising_model}

In the following, we consider the pure LRIM with vanishing transverse field $h=0$. Here it is a priori not clear what classical state is the ground state as a function of~$\alpha$. In the following we want to clarify the nature of the ground state for the YC($n$) cylinders and its ground-state energy per site, which is then used to set up the low-field expansion.

% Figure stripes
%%%%%%%%%%%%%%%%%%%%%%%%%%%%%%%%%%%%%%%%%%%
\begin{figure}
	\centering
	\includegraphics{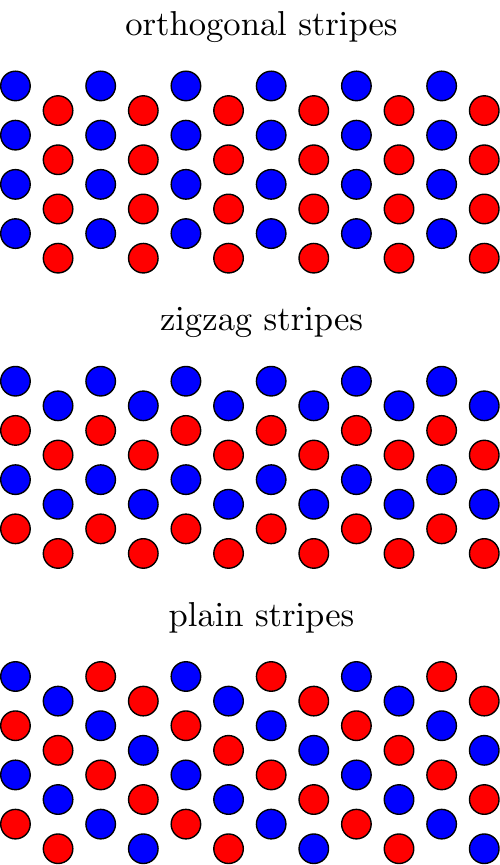}
	\caption{Illustration of orthogonal-, zigzag-, and plain-stripe states in real space on a YC($4$) lattice. Red and blue circles denote spins pointing in opposite directions. The infinite extension of the cylinder is in horizontal direction.}
	\label{fig:stripes}
\end{figure}
%%%%%%%%%%%%%%%%%%%%%%%%%%%%%%%%%%%%%%%%%%%

By studying the energy of the nearest-neighbor ground states on finite cylinders with $N\approx40$ spins for $\alpha<\infty$, we observe that the relevant states for the LRIM consist only of \emph{non-flippable} plaquettes in the quantum dimer language on the dual lattice. These states are symmetry-broken, therefore gapped and stable against quantum fluctuations introduced by small transverse magnetic fields. We name the three relevant occurring order patterns \emph{orthogonal} [$\vec{k}=(\pi,0)^T$], \emph{plain} [$\vec{k}=(\pi,\pi)^T$], and \mbox{\emph{zigzag} [$\vec{k}=(\pi/2,\pi)^T$]} \emph{stripes}, which are illustrated in Fig.~\ref{fig:stripes}. 

%
% Figures: e0 - classical stripes YC4/YC6
%%%%%%%%%%%%%%%%%%%%%%%%%%%%%%%%%%%%%%%%%%%%%%%%%%%
\begin{figure}
	\centering
	\includegraphics{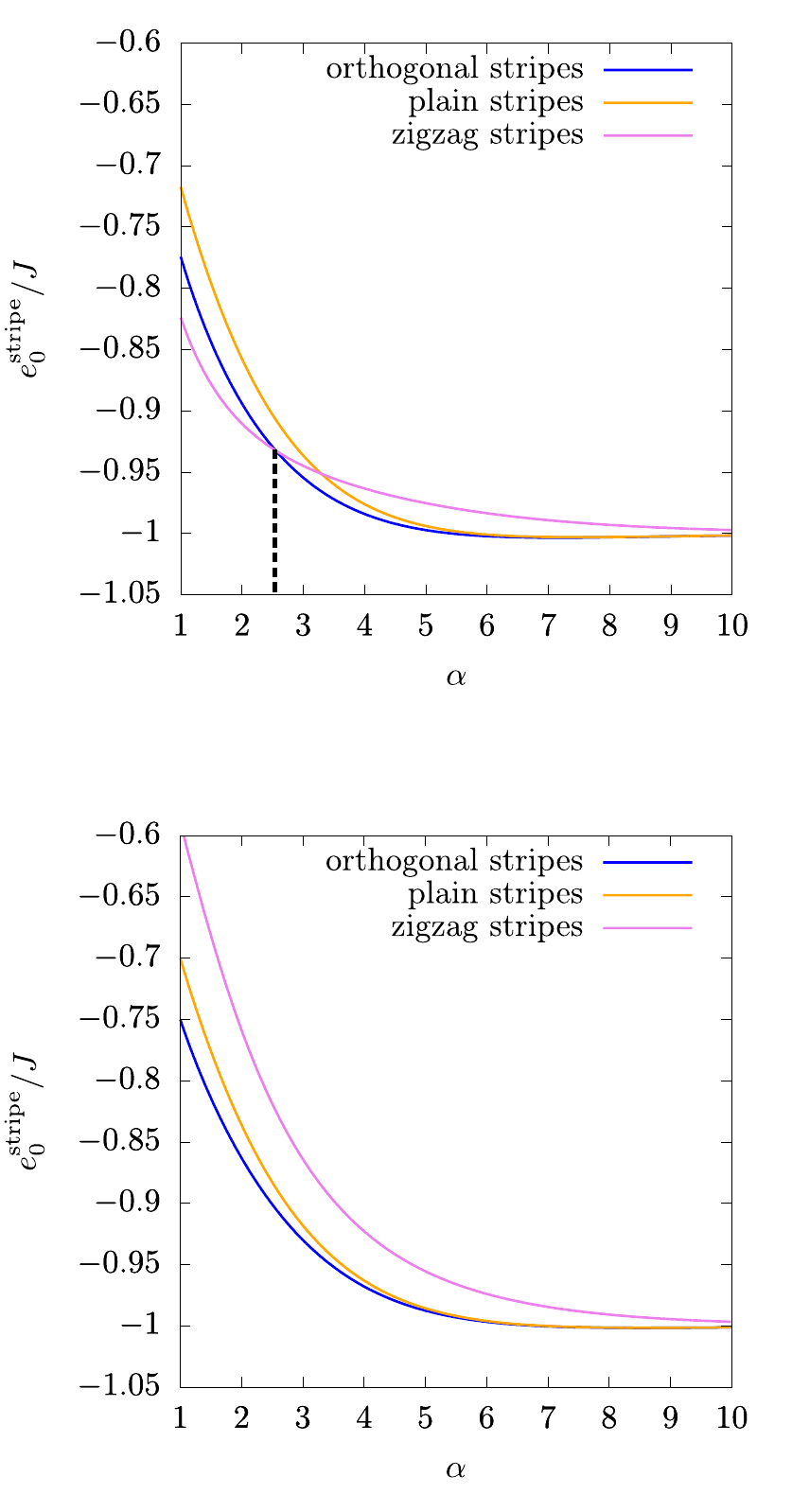}
	\caption{Energy per site $e_0^{\rm stripe}$ for the considered stripe patterns evaluated on the YC($4$) lattice for $N=4\cdot1000$ (upper panel) and on the YC($6$) lattice for $N=6\cdot1000$ (lower panel) with periodic boundary conditions. The dotted vertical line in the upper panel indicates the first-order phase transition $\alpha_{\rm c}$ between zigzag ($\alpha<\alpha_{\rm c}$) and orthogonal ($\alpha>\alpha_{\rm c}$) stripes for the YC($4$) cylinder. For the YC($6$) orthogonal stripes are realized for all $\alpha$.}
	\label{fig:zero_field_46}
\end{figure}
%%%%%%%%%%%%%%%%%%%%%%%%%%%%%%%%%%%%%%%%%%%%%%%%%%%

In the orthogonal-stripe configuration spins with the same orientation order orthogonally to the direction of the cylinder, which leads to a two-fold degeneracy which results from the $\mathbb{Z}_2$ spin-flip symmetry of the state. In the plain-stripe state spins of the same orientation align in plain chains winding around the cylinder in direction of infinite extension. This pattern has a $\mathbb{Z}_2\times\mathbb{Z}_2$ symmetry, which results from a spin-flip symmetry and a decoupling of the state into two sublattices. In the 2D limit orthogonal and plain stripes are degenerate with a $\mathbb{Z}_2\times\mathbb{Z}_3$ symmetry due to spin flips and the threefold rotational symmetry of the lattice. The absence of this rotational symmetry for the YC($n$) cylinders leads to the energetic separation of orthogonal and plain stripes, where orthogonal stripes have lower energies for all decay exponents $\alpha$ on the YC($4$) and YC($6$) lattice. In fact, the preference of orthogonal stripes is seen for all studied YC$(n)$ cylinders with even $n$.

The third identified ground-state pattern are the zigzag stripes where spins of the same orientation align in a zigzag shape in cylinder direction. This results in a four-fold degeneracy ($\mathbb{Z}_2\times\mathbb{Z}_2$) due to spin-flip symmetry and a decoupling into two sublattices. On the YC$(n)$ cylinders with $n=4s\ (s \in \mathbb{N})$ it is actually possible to rotate the zigzag stripes by $2\pi/3$ and to remain in the subspace of states with only non-flippable plaquettes. These rotated zigzag stripes are always energetically less beneficial than the zigzag stripes with an alignment in infinite direction and we will not consider them further.

At small $\alpha$ zigzag stripes are energetically lower compared to the orthogonal stripes for the YC($4$), YC($8$), and YC($12$) cylinder (see Fig.~\ref{fig:zero_field_46}). Consequently, there must be a first-order phase transition between these two stripe phases for these cylinders and we can determine the associated critical $\alpha_{\text{c}}$ which are listed in Tab.~\ref{tab:krit_alpha}.
\begin{table}[!ht]
	\centering
	\begin{tabular}{c||c|c|c}
		$n$ &	$4$ &	$8$ &	$12$ \\ \hline
		$\alpha_{\text{c}}$ & 2.55(1) & 1.41(1) & 1.13(1)
	\end{tabular}
	\caption{Critical decay exponents $\alpha_c$ for a phase transition between orthogonal ($\alpha>\alpha_{\rm c}$) and zigzag ($\alpha<\alpha_{\rm c}$) stripes on the YC($n$) lattices with $n\in\{4,8,12\}$. The denoted lattices are the ones where the transition occurs in the considered range of $\alpha>1$ in the framework of our numerical real-space implementation.}
	\label{tab:krit_alpha}
\end{table}
The favoring of these stripe patterns can be explained by looking at the dominant further-neighbor Ising interactions, which are contributing differently to the energy, depending on the type of stripe under consideration. For zigzag stripes the second-nearest neighbors are contributing less beneficial than for the orthogonal stripes, but the third-nearest neighbors are lowering the energy more than they do for the orthogonal stripes. Together with the periodicity of the YC($n$) cylinder this leads to a favoring of the zigzag stripes at low $\alpha$ for the lattices described above, since then the long-range Ising interactions come more and more into play. Such stripe patterns have also been found by Smerald \etal~for a truncated long-range Ising interaction \cite{Smerald2016}. For a classical Hamiltonian with arbitrary tunable antiferromagnetic nearest and next-nearest neighbor interaction it is a well established result, that on the triangular 2D lattice $\mathbb{Z}_2\times\mathbb{Z}_3$ plain stripes realize the ground state of the system \cite{Metcalf1974, Korshunov2005}. On any YC$(n)$ lattice, this degeneracy of plain and orthogonal stripes is lifted. Compared to the plain-stripe order, the $(n+1)$-nearest neighbors of the orthogonal stripes are missing two ferromagnetic interactions for each site along the cylinder ring due to the finite cylinder extension. Instead two additional antiferromagnetic interactions per site are present which leads in total to a lower energy of orthogonal stripes. A second consequence of the cylinder geometry is the relevance of a zigzag-striped order. Further, we note that these three stripe patterns are part of the $\alpha=0$ and $\alpha=\infty$ ground-state space. Using the $\alpha=0$ limit of the LRIM with the perturbation described in Eq.~\eqref{eq:smallalphalogperturb} confirms the above findings. To this end, we considered all $\alpha=0$ ground states on a cluster of $N \approx 40$ spins which leads to the same ground-state space of non-flippable plaquettes. Evaluating the energy for the three relevant stripe patterns on large clusters $N\approx10000$, we find the same stripe-ordered ground states as calculated directly from the full LRIM.

We therefore find that the physical behavior of the LRIM is different for the two families of cylinders with $n=4s\ (s\in \mathbb{N})$ and $n=4s+2 \ (s\in \mathbb{N})$. For $n=4s+2$ orthogonal stripes are realized for all studied $\alpha$, while in the other case orthogonal stripes become unstable towards a zigzag-stripe order for small $\alpha$. This can be clearly seen in the energy evolution of the different stripe structures as a function of $\alpha$, which is shown for the smallest member YC($4$) and YC($6$) of both families in Fig.~\ref{fig:zero_field_46}, where we have studied finite cylinders with  $N($YC$(n))=n\cdot1000$ spins using periodic boundary conditions. In the following, we focus on these two cylinders and study the ground-state phase diagram of the full LRTFIM. 

% section pure_long_range_ising_model (end)

\section{Phase diagrams} % (fold)
\label{sec:phase_diagrams}

As described above, we performed series expansions to extract the ground-state energies of the $x$-polarized, stripe, and clock-ordered phases for the LRTFIM on the YC($4$) and YC($6$) cylinder. Quantum phase transitions between two of these phases can then be located by determining crossing points between these energies for fixed $\alpha$. Note that this approach is not sensitive enough to extract the nature of the phase transition, although strong first-order phase transitions are expected to be located with highest precision. Furthermore, we study the breakdown of the $x$-polarized phase by investigating the behavior of the one-particle excitation energies for specific momenta, which are associated with the translational symmetry of the clock- or stripe-ordered phases. A continuous gap-closing of a certain mode signals a continuous quantum phase transition between the $x$-polarized phase and the associated ordered phase. We start our discussion with the YC($6$) cylinder, where the clock order is identical to the one known for the LRTFIM on the 2D triangular lattice and which has already been studied by Saadatmand et al.\cite{Saadatmand2018}. Afterwards, we turn our attention to the YC($4$) cylinder on which this clock order is frustrated and the ground-state phase diagram is even richer.
%Finally, we discuss the generic presence of an intermediate critical phase on YC($n$) cylinders, which is situated in the phase diagram between the $x$-polarized and the clock-ordered phase. The expected KT-transitions out of this intermediate phase cannot be captured by our techniques and a detailed characterization certainly needs further studies. 

\subsection{YC$(6)$} % (fold)
\label{sub:YC6}

%
% Figure: PD YC6
%%%%%%%%%%%%%%%%%%%%%%%%%%%%%%%%%%%%%%%%%%%%%%%%%%%
\begin{figure}
	\centering
	\includegraphics{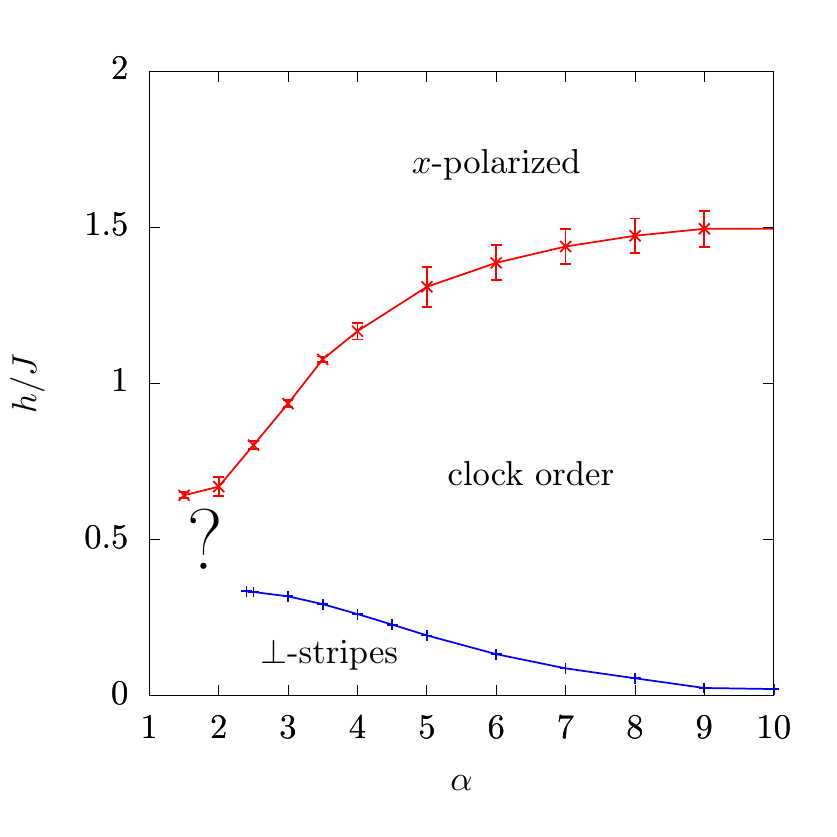}
	\caption{Ground-state phase diagram calculated for the YC($6$) cylinder using series expansions. The red line is determined by the closing of the one-QP gap \mbox{$\Delta_{\rm 1QP}$} at momentum \mbox{$(2\pi/3,-2\pi/3)^T$} with the standard deviation of the Pad{\'e} approximants up to tenth order. The blue line is determined by the energy intersection between the clock-order energy $e_0^{\rm clock}$ and the stripe energies in sixth order. The question mark indicates the region of the phase diagram where the used methods break down.}
	\label{fig:phasediagramYC6}
\end{figure}
\begin{figure}
	\centering
	\includegraphics{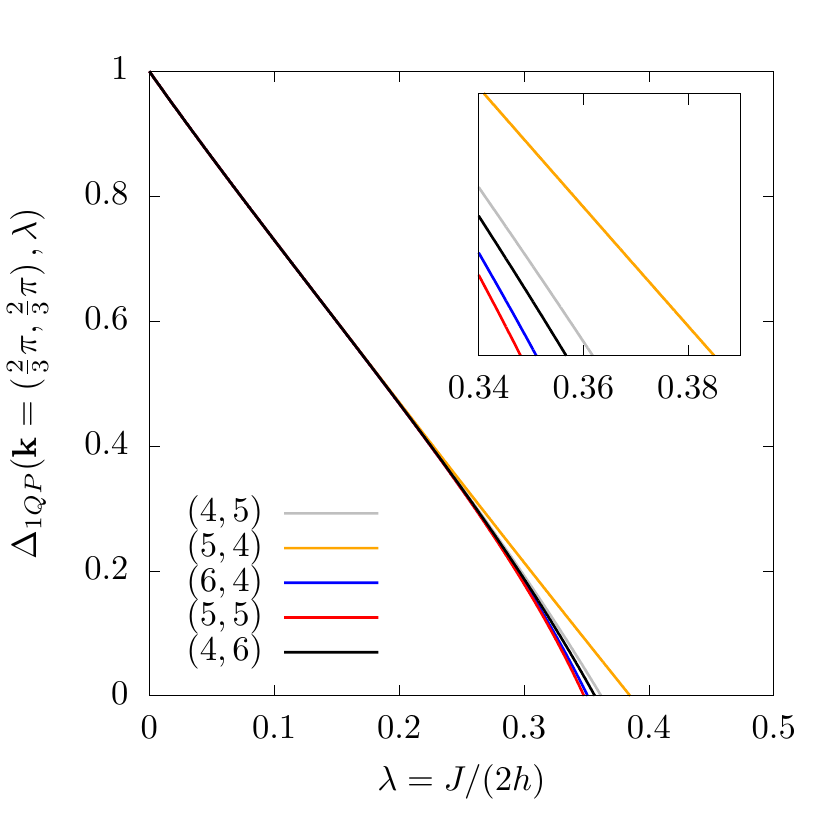}
	\caption{Exemplary depiction of the phase transition points for $\alpha=6$ between the x-polarized phase and the clock ordered phase for the YC$(6)$ cylinder. The figure displays the one-QP gap \mbox{$\Delta_{\rm 1QP}$} at momentum \mbox{$(2\pi/3,-2\pi/3)^T$} as a function of $\lambda=h/2J$ using the order nine and ten Pad\'{e} extrapolants with no poles before the gap closing. Depicted Pad\'{e} extrapolants are $(4,5)$, $(5,4)$, $(5,5)$, $(6,4)$ and $(4,6)$.}
	\label{fig:a6pcut}
\end{figure}

\begin{figure}
	\centering
	\includegraphics{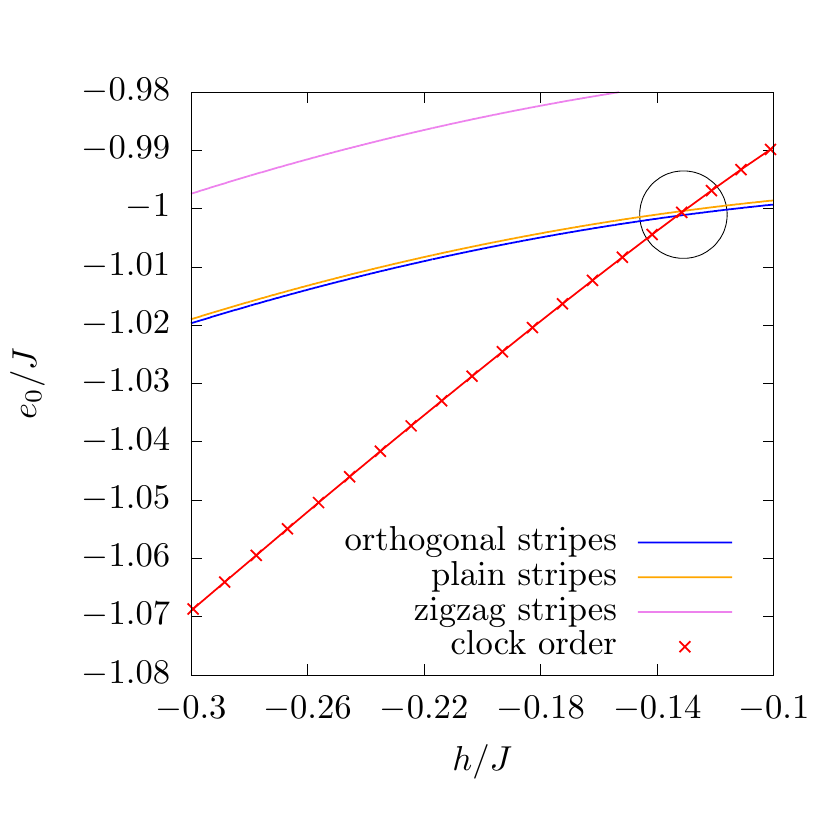}
	\caption{Exemplary depiction of the phase-transition points for $\alpha=6$ between the orthogonal-stripe phase and the clock-ordered phase for the YC$(6)$ cylinder. The figure shows the ground-state energies of the stripe and clock-order phases as a function of $h/J$. Red crosses denote calculated points in the parameter space of $h$ and $\alpha=6$ for the clock-order energy, the blue line denotes the orthogonal-stripe energy, the yellow line denotes the plain-stripe energy, and the violet line represents the zigzag-stripe energy. The crossing between the red symbols and the blue line indicates the first-order phase transition.}
	\label{fig:a6stripe}
\end{figure}

%%%%%%%%%%%%%%%%%%%%%%%%%%%%%%%%%%%%%%%%%%%%%%%%%%%
The obtained ground-state phase diagram for the \mbox{LRTFIM} on the YC$(6)$ cylinder is shown in Fig.~\ref{fig:phasediagramYC6}. It displays the $x$-polarized phase, the clock order, and the orthogonal-stripe phase. The quantum phase transition between the $x$-polarized phase and the clock order is located by investigating the one-particle excitation energies of the $x$-polarized phase using the high-field expansion. For a non-first-order phase transition one expects that the one-particle gap of the $x$-polarized phase closes at the quantum-critical point and has a momentum $\vec k=(2\pi/3,-2\pi/3)^T$, which is associated with the clock order. Here we locate such a gap-closing quantum-critical point by applying Pad{\'e} extrapolations on the bare order 10 series (see Fig.~\ref{fig:a6pcut}) and we quantify the uncertainty of this extrapolation scheme by the standard deviation of different extrapolations shown as error bars in Fig.~\ref{fig:phasediagramYC6}. This gap-closing we can track up to decay exponents $\alpha=1.5$. The calculated phase-transition points for $\alpha\gtrsim 2.4$ are within error bars in good agreement with the numerical findings by Saadatmand \etal \cite{Saadatmand2018}. Specifically, for the NNTFIM $\alpha\rightarrow\infty$, the pCUT high-field calculation yields a gap closing at a transverse field $h=1.54(7)\,J$, which has to be compared to $h=1.5(1)\,J$ determined numerically by investigating the order parameter for the clock order \cite{Saadatmand2018}.

The second type of phase transition present in the phase diagram is between the clock order and the orthogonal-stripe phase (see Fig.~\ref{fig:a6stripe} and blue line in Fig.~\ref{fig:phasediagramYC6}). Since both phases break a different type of discrete translational lattice symmetry, this transition is first order. It can therefore be located by determining the level crossing $e_0^{\rm clock}=e_0^{\rm stripe}$ for a given value of $\alpha$. As the ground-state energy $e_0^{\rm clock}$ of the clock-ordered state is evaluated on a finite cluster of six rings and the long-range interactions are included perturbatively, one has to be aware that the blue line in Fig.~\ref{fig:phasediagramYC6} is certainly not quantitative for small values of $\alpha$. In fact, we expect the phase transition to occur at higher transverse fields $h/J$ for small $\alpha$, because a better treatment of the long-range interactions would result in an increased ground-state energy $e_0^{\rm clock}$ so that the orthogonal-stripe phase is enlarged with respect to the clock order. In contrast, we (slightly) overestimate $e_0^{\rm clock}$ for large $\alpha$ due to the finite cluster extension and the approximate treatment of the field-induced quantum fluctuations. It is therefore plausible that for $\alpha \lesssim 2.4$ no clock order is present anymore in the phase diagram as suggested by Saadatmand \etal\cite{Saadatmand2018} and there is a direct phase transition between the $x$-polarized and the orthogonal-stripe phase. Another scenario is the presence of an intermediate (gapless) phase as we discuss in Sect.~\ref{sec:conclusion}. We stress again that orthogonal stripes are the true ground states of the zero-field LRIM and no zigzag stripes are realized in the LRTFIM as found by Ref.~\onlinecite{Saadatmand2018}.

If there is a direct continuous phase transition between the orthogonal-stripe and the $x$-polarized phase, one expects that the high-field gap in the $x$-polarized phase closes at the critical point and is located at the associated momentum of the orthogonal-stripe phase $\vec k=(\pi,0)^T$ for $\alpha \lesssim 2.4$. To study the transition to the orthogonal-stripe phase, we therefore evaluated the one-QP energy in the high-field limit at momentum $\vec k=(\pi,0)^T$ using Pad\'{e} extrapolations. Interestingly, no closing of the gap could be observed so that we cannot confirm a direct phase transition between the $x$-polarized and the orthogonal-stripe phase using series expansion methods. So either the situation is similar to the case of the 2D triangular lattice, where the direct phase transition is known to be generically first order due to the $\mathbb{Z}_2\cross \mathbb{Z}_3$ symmetry of the stripe order, or the phase diagram contains an intermediate phase which we elaborate on further in the conclusion. 

% subsection YC6 (end)

\subsection{YC$(4)$} % (fold)
\label{sub:YC4}

%
% Figure: PD YC4
%%%%%%%%%%%%%%%%%%%%%%%%%%%%%%%%%%%%%%%%%%%%%%%%%%%
\begin{figure}
	\centering
	\includegraphics{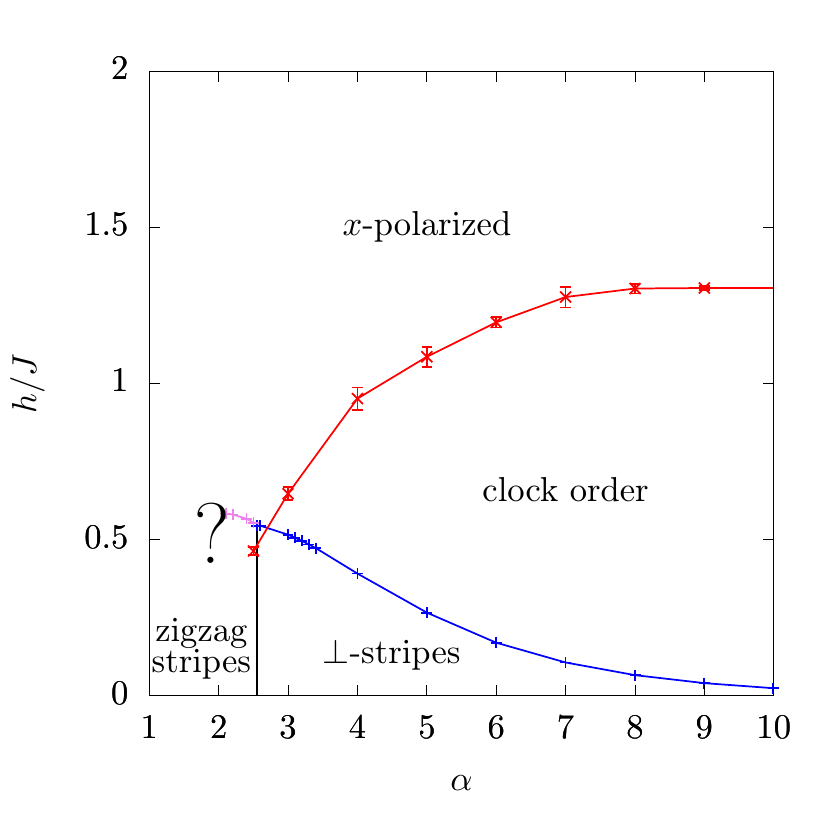}
	\caption{Ground-state phase diagram calculated for the YC($4$) cylinder using series expansions. The red line is determined by the closing of the one-QP gap with the standard deviation of the Pad{\'e} approximants up to order ten. The blue line is determined by the energy intersection between the clock-order energy $e_0^{\rm clock}$ and the orthogonal ($\perp$) stripe energy in sixth order. The violet line denotes the intersection between the energy of the quantum dimer model in third oder and the zigzag stripe energy in sixth order. The almost vertical black line represents the first-order phase transition line between the zigzag and orthogonal stripes, which is determined by comparing the ground-state energies of both stripe phases. The question mark indicates the region of the phase diagram where the used methods break down.}
	\label{fig:phasediagram_YC4}
      \end{figure}
%%%%%%%%%%%%%%%%%%%%%%%%%%%%%%%%%%%%%%%%%%%%%%%%%%%    

Let us turn to the ground-state phase diagram for the LRTFIM on the YC$(4)$ cylinder, which is shown in Fig.~\ref{fig:phasediagram_YC4}. The phase boundaries were calculated analogously to the YC($6$) lattice, where the $x$-polarized- to clock-order transition was determined by Pad\'{e} extrapolations of the perturbative expansion from the high-field limit and the other transition lines were derived from the crossing of the respective ground-state energies.

Besides the $x$-polarized phase a clock-ordered phase with a different momentum $\vec k = (5\pi/4,\pi/2)^T$ from the YC($6$) lattice arises. Additional to the orthogonal-stripe order with $\vec k=(\pi,0)^T$ at finite large $\alpha$ already discussed for the YC($6$) lattice, we find zigzag stripes with \mbox{$\vec k=(\pi/2,\pi)^T$} for small fields. The first-order phase transition from the orthogonal to a zigzag order for small fields occurs at $\alpha \approx 2.55(1)$. Interestingly, the phase transition between these two stripe orders is almost independent of $h/J$ so that a nearly vertical phase transition line results (see black line in Fig.~\ref{fig:phasediagram_YC4}). Further, for small $\alpha$, one might predict a direct phase transition between the $x$-polarized phase realized at high field strength and the zigzag stripes at low field strength in similarity to the YC($6$) cylinder. Pad\'{e} extrapolations of the one-QP gap with $\vec k=(\pi/2,\pi)^T$ again do not point towards a continuous phase transition signaled by a gap closing in this $\alpha$-regime. We therefore expect that the physical situation is similar to the YC($6$) cylinder as discussed below.

In the calculated phase diagram the intersection between the clock-order energy and the different stripe orders is calculated for all $\alpha$ values for which the clock-order expansion is available. Even though the clock-order is not expected to be the ground state for small $\alpha$, the calculated line gives a reference point for the extension of the stripe-ordered phases. As for the YC$(6)$ cylinder, we have to stress that the calculated energy for the clock order is underestimated, because the long-range interactions is cutted due to the evaluation on finite clusters. This implies that the transition between stripes and clock order occurs at higher transverse fields for small decay exponents ($\alpha \lessapprox 3.5$).

% subsection YC4 (end)

% section phase_diagrams (end)

\section{Conclusions}
\label{sec:conclusion}
We have investigated the ground-state phase diagram of the LRTFIM on triangular-lattice cylinders using various approaches. The physical behavior of the classical LRIM is different for the two families of cylinders with $n=4s\ (s\in \mathbb{N})$ and $n=4s+2 \ (s\in \mathbb{N})$. For $n=4s+2$ orthogonal stripes are realized for all studied $\alpha$, while in the other case orthogonal stripes become unstable towards a zigzag-stripe order for small $\alpha$. For the YC($6$) cylinder, our results are therefore distinct from the zigzag stripes obtained numerically \cite{Saadatmand2018}, which is most likely due to the chosen unit cell in the iDMRG approach \cite{McCulloch_Private}. The full quantum phase diagram of the LRTFIM on the YC($4$) and YC($6$) cylinder contains at least three different types of gapped quantum phases. An $x$-polarized paramagnetic high-field phase, stripe phases triggered by the long-range Ising interaction, as well as clock-ordered phases being stabilized via an order-by-disorder mechanism about the highly-degenerate classical spin liquid of the nearest-neighbor Ising model. The extension of these phases in the parameter space of the LRFTIM has been located approximately by a variety of different perturbative expansions. In contrast to the high- and low-field high-order expansion, the obtained ground-state energy of the clock-ordered phase is the least accurate due to the low order three of the effective Hamiltonian and due to the finite cluster size. The obtained ground-state phase diagrams are valid as long as one assumes that no other phase is present. However, this is not obvious, which leads to the following two points which deserve further investigations:

First, we turn our attention to the nearest-neighbor TFIM on the YC($6$) cylinder, which realizes the same kind of clock order as the TFIM on the two-dimensional triangular lattice. For the 2D triangular lattice the quantum phase transition between the $x$-polarized and the clock-ordered phase is a continuous second-order transition which falls into the $(2+1)$D-XY universality class \cite{Moessner2001,Moessner2003,Powalski2013}. This follows from a mapping of the TFIM onto a classical XY-model in three dimensions \cite{Blankschtein1984}. By applying the same kind of quantum to classical mapping on our quasi one-dimensional cylinders, one would expect that the transition falls into the $(1+1)$D-XY universality class, which is known to be the archetype of an infinite-order Kosterlitz-Thouless phase transition \cite{Kosterlitz1973}. However, since the Kosterlitz-Thouless phase transition involves one phase with critical (algebraically-decaying) correlations and both, the $x$-polarized and the clock-ordered phase, are gapped, this implies the existence of a gapless intermediate phase in the ground-state phase diagram of the nearest-neighbor TFIM on the YC(6) cylinder so that there are two Kosterlitz-Thouless transitions out of this intermediate phase. For the corresponding classical phase transitions in the NNTFIM on the 2D triangular lattice as a function of temperature, such an intermediate phase as well as the associated Kosterlitz-Thouless transitions are well established theoretically \cite{Moessner2000,Moessner2003,Wang2017} and confirmed experimentally recently in the Ising-type triangular antiferromagnet TmMgGaO$_4$ \cite{Li2019a}. Clearly, this intermediate phase in the YC($6$) cylinder will also extend in a finite $\alpha$-window in the phase diagram of the LRFTIM and, by similarity, one would expect a similar phase also for the YC(4) cylinder. Let us note that the detection of a Kosterlitz-Thouless transition is not possible with our high-field expansion due to the non-analytic behavior of the gap close to such a phase transition. Further, we find it interesting that iDMRG calculations based on translational invariant states are in good agreement with our findings with respect to the phase transition line between the $x$-polarized and clock-ordered phase as a function of $\alpha$ \cite{Saadatmand2018}. It is likely that both approaches are not sensitive enough to pinpoint the intermediate phase on the quasi one-dimensional YC(6) cylinder, but rather yield a good estimation for the transition line of the corresponding two-dimensional system on the triangular lattice. In any case, the existence and nature of the intermediate phase has to be clarified in the future.

Second, it is not clear how the ground-state phase diagram looks for smaller values of $\alpha$, when the clock order (and potentially the just-discussed intermediate phase, are not realized anymore. The numerical work of Ref.~\onlinecite{Saadatmand2018} on the YC(6) cylinder suggests a direct second-order phase transition between the $x$-polarized and a stripe-ordered phase. We stress again that our investigation yields clearly a different ordering pattern for the stripes, namely orthogonal stripes. Extrapolations of the one-particle high-field gap with the corresponding stripe-momentum give no evidence for a gap-closing (second-order) phase transition. So this phase transition might be either (weakly) first order as for the LRFTIM on the triangular lattice \cite{Korshunov2005,Smerald2016,Fey2019} or, again, an intermediate phase could be present between the $x$-polarized and the stripe phase which prevents a controlled extrapolation of the gap. An indication for the latter scenario might be the presence of an intermediate classical spin liquid as a function of temperature for a deformed classical Ising model with dipolar interactions on the triangular lattice \cite{Smerald2018}.

Overall, the interplay of geometric frustration and long-range interactions in low-dimensional quantum magnets displays a variety of interesting quantum phenomena which certainly need further investigations in the future.

% Acknowledgements
%%%%%%%%%%%%%%%%%%%%%%%%%%%%%%%%%%%%%%%%%%%%%%%%%%%%%%%%%%%%%%%%%%%%%%%%%%%%%%%%%%%%%%%%%%%%

\section{Acknowledgments}
We thank Ian McCulloch and Andrew Smerald for fruitful discussions. We gratefully acknowledge the compute resources and support provided by the HPC group of the Erlangen Regional Computing Center (RRZE).

\bibliography{bibliography}

%merlin.mbs apsrev4-1.bst 2010-07-25 4.21a (PWD, AO, DPC) hacked
%Control: key (0)
%Control: author (72) initials jnrlst
%Control: editor formatted (1) identically to author
%Control: production of article title (-1) disabled
%Control: page (0) single
%Control: year (1) truncated
%Control: production of eprint (0) enabled
\begin{thebibliography}{39}%
\makeatletter
\providecommand \@ifxundefined [1]{%
 \@ifx{#1\undefined}
}%
\providecommand \@ifnum [1]{%
 \ifnum #1\expandafter \@firstoftwo
 \else \expandafter \@secondoftwo
 \fi
}%
\providecommand \@ifx [1]{%
 \ifx #1\expandafter \@firstoftwo
 \else \expandafter \@secondoftwo
 \fi
}%
\providecommand \natexlab [1]{#1}%
\providecommand \enquote  [1]{``#1''}%
\providecommand \bibnamefont  [1]{#1}%
\providecommand \bibfnamefont [1]{#1}%
\providecommand \citenamefont [1]{#1}%
\providecommand \href@noop [0]{\@secondoftwo}%
\providecommand \href [0]{\begingroup \@sanitize@url \@href}%
\providecommand \@href[1]{\@@startlink{#1}\@@href}%
\providecommand \@@href[1]{\endgroup#1\@@endlink}%
\providecommand \@sanitize@url [0]{\catcode `\\12\catcode `\$12\catcode
  `\&12\catcode `\#12\catcode `\^12\catcode `\_12\catcode `\%12\relax}%
\providecommand \@@startlink[1]{}%
\providecommand \@@endlink[0]{}%
\providecommand \url  [0]{\begingroup\@sanitize@url \@url }%
\providecommand \@url [1]{\endgroup\@href {#1}{\urlprefix }}%
\providecommand \urlprefix  [0]{URL }%
\providecommand \Eprint [0]{\href }%
\providecommand \doibase [0]{http://dx.doi.org/}%
\providecommand \selectlanguage [0]{\@gobble}%
\providecommand \bibinfo  [0]{\@secondoftwo}%
\providecommand \bibfield  [0]{\@secondoftwo}%
\providecommand \translation [1]{[#1]}%
\providecommand \BibitemOpen [0]{}%
\providecommand \bibitemStop [0]{}%
\providecommand \bibitemNoStop [0]{.\EOS\space}%
\providecommand \EOS [0]{\spacefactor3000\relax}%
\providecommand \BibitemShut  [1]{\csname bibitem#1\endcsname}%
\let\auto@bib@innerbib\@empty
%</preamble>
\bibitem [{\citenamefont {Kitaev}(2006)}]{Kitaev2006}%
  \BibitemOpen
  \bibfield  {author} {\bibinfo {author} {\bibfnamefont {A.}~\bibnamefont
  {Kitaev}},\ }\href {\doibase 10.1016/j.aop.2005.10.005} {\bibfield  {journal}
  {\bibinfo  {journal} {Annals of Physics}\ }\textbf {\bibinfo {volume}
  {321}},\ \bibinfo {pages} {2 } (\bibinfo {year} {2006})},\ \bibinfo {note}
  {january Special Issue}\BibitemShut {NoStop}%
\bibitem [{\citenamefont {Castelnovo}\ \emph {et~al.}(2008)\citenamefont
  {Castelnovo}, \citenamefont {Moessner},\ and\ \citenamefont
  {Sondhi}}]{Castelnovo2008}%
  \BibitemOpen
  \bibfield  {author} {\bibinfo {author} {\bibfnamefont {C.}~\bibnamefont
  {Castelnovo}}, \bibinfo {author} {\bibfnamefont {R.}~\bibnamefont
  {Moessner}}, \ and\ \bibinfo {author} {\bibfnamefont {S.~L.}\ \bibnamefont
  {Sondhi}},\ }\href {\doibase 10.1038/nature06433} {\bibfield  {journal}
  {\bibinfo  {journal} {Nature}\ }\textbf {\bibinfo {volume} {451}},\ \bibinfo
  {pages} {42} (\bibinfo {year} {2008})}\BibitemShut {NoStop}%
\bibitem [{\citenamefont {Nagle}\ and\ \citenamefont
  {Bonner}(1970)}]{Nagle1970}%
  \BibitemOpen
  \bibfield  {author} {\bibinfo {author} {\bibfnamefont {J.~F.}\ \bibnamefont
  {Nagle}}\ and\ \bibinfo {author} {\bibfnamefont {J.~C.}\ \bibnamefont
  {Bonner}},\ }\href {\doibase 10.1088/0022-3719/3/2/017} {\bibfield  {journal}
  {\bibinfo  {journal} {Journal of Physics C: Solid State Physics}\ }\textbf
  {\bibinfo {volume} {3}},\ \bibinfo {pages} {352} (\bibinfo {year}
  {1970})}\BibitemShut {NoStop}%
\bibitem [{\citenamefont {Fisher}\ \emph {et~al.}(1972)\citenamefont {Fisher},
  \citenamefont {Ma},\ and\ \citenamefont {Nickel}}]{Fisher1972}%
  \BibitemOpen
  \bibfield  {author} {\bibinfo {author} {\bibfnamefont {M.~E.}\ \bibnamefont
  {Fisher}}, \bibinfo {author} {\bibfnamefont {S.-k.}\ \bibnamefont {Ma}}, \
  and\ \bibinfo {author} {\bibfnamefont {B.~G.}\ \bibnamefont {Nickel}},\
  }\href {\doibase 10.1103/PhysRevLett.29.917} {\bibfield  {journal} {\bibinfo
  {journal} {Phys. Rev. Lett.}\ }\textbf {\bibinfo {volume} {29}},\ \bibinfo
  {pages} {917} (\bibinfo {year} {1972})}\BibitemShut {NoStop}%
\bibitem [{\citenamefont {Dutta}\ and\ \citenamefont
  {Bhattacharjee}(2001)}]{Dutta2001}%
  \BibitemOpen
  \bibfield  {author} {\bibinfo {author} {\bibfnamefont {A.}~\bibnamefont
  {Dutta}}\ and\ \bibinfo {author} {\bibfnamefont {J.~K.}\ \bibnamefont
  {Bhattacharjee}},\ }\href {\doibase 10.1103/PhysRevB.64.184106} {\bibfield
  {journal} {\bibinfo  {journal} {Phys. Rev. B}\ }\textbf {\bibinfo {volume}
  {64}},\ \bibinfo {pages} {184106} (\bibinfo {year} {2001})}\BibitemShut
  {NoStop}%
\bibitem [{\citenamefont {Knap}\ \emph {et~al.}(2013)\citenamefont {Knap},
  \citenamefont {Kantian}, \citenamefont {Giamarchi}, \citenamefont {Bloch},
  \citenamefont {Lukin},\ and\ \citenamefont {Demler}}]{Knap2013}%
  \BibitemOpen
  \bibfield  {author} {\bibinfo {author} {\bibfnamefont {M.}~\bibnamefont
  {Knap}}, \bibinfo {author} {\bibfnamefont {A.}~\bibnamefont {Kantian}},
  \bibinfo {author} {\bibfnamefont {T.}~\bibnamefont {Giamarchi}}, \bibinfo
  {author} {\bibfnamefont {I.}~\bibnamefont {Bloch}}, \bibinfo {author}
  {\bibfnamefont {M.~D.}\ \bibnamefont {Lukin}}, \ and\ \bibinfo {author}
  {\bibfnamefont {E.}~\bibnamefont {Demler}},\ }\href {\doibase
  10.1103/PhysRevLett.111.147205} {\bibfield  {journal} {\bibinfo  {journal}
  {Phys. Rev. Lett.}\ }\textbf {\bibinfo {volume} {111}},\ \bibinfo {pages}
  {147205} (\bibinfo {year} {2013})}\BibitemShut {NoStop}%
\bibitem [{\citenamefont {Fey}\ and\ \citenamefont {Schmidt}(2016)}]{Fey2016}%
  \BibitemOpen
  \bibfield  {author} {\bibinfo {author} {\bibfnamefont {S.}~\bibnamefont
  {Fey}}\ and\ \bibinfo {author} {\bibfnamefont {K.~P.}\ \bibnamefont
  {Schmidt}},\ }\href {\doibase 10.1103/PhysRevB.94.075156} {\bibfield
  {journal} {\bibinfo  {journal} {Phys. Rev. B}\ }\textbf {\bibinfo {volume}
  {94}},\ \bibinfo {pages} {075156} (\bibinfo {year} {2016})}\BibitemShut
  {NoStop}%
\bibitem [{\citenamefont {Defenu}\ \emph {et~al.}(2017)\citenamefont {Defenu},
  \citenamefont {Trombettoni},\ and\ \citenamefont {Ruffo}}]{Defenu2017}%
  \BibitemOpen
  \bibfield  {author} {\bibinfo {author} {\bibfnamefont {N.}~\bibnamefont
  {Defenu}}, \bibinfo {author} {\bibfnamefont {A.}~\bibnamefont {Trombettoni}},
  \ and\ \bibinfo {author} {\bibfnamefont {S.}~\bibnamefont {Ruffo}},\ }\href
  {\doibase 10.1103/PhysRevB.96.104432} {\bibfield  {journal} {\bibinfo
  {journal} {Physical Review B}\ }\textbf {\bibinfo {volume} {96}},\ \bibinfo
  {pages} {104432} (\bibinfo {year} {2017})}\BibitemShut {NoStop}%
\bibitem [{\citenamefont {Fey}\ \emph {et~al.}(2019)\citenamefont {Fey},
  \citenamefont {Kapfer},\ and\ \citenamefont {Schmidt}}]{Fey2019}%
  \BibitemOpen
  \bibfield  {author} {\bibinfo {author} {\bibfnamefont {S.}~\bibnamefont
  {Fey}}, \bibinfo {author} {\bibfnamefont {S.~C.}\ \bibnamefont {Kapfer}}, \
  and\ \bibinfo {author} {\bibfnamefont {K.~P.}\ \bibnamefont {Schmidt}},\
  }\href {\doibase 10.1103/PhysRevLett.122.017203} {\bibfield  {journal}
  {\bibinfo  {journal} {Phys. Rev. Lett.}\ }\textbf {\bibinfo {volume} {122}},\
  \bibinfo {pages} {017203} (\bibinfo {year} {2019})}\BibitemShut {NoStop}%
\bibitem [{\citenamefont {Schau{\ss}}\ \emph {et~al.}(2012)\citenamefont
  {Schau{\ss}}, \citenamefont {Cheneau}, \citenamefont {Endres}, \citenamefont
  {Fukuhara}, \citenamefont {Hild}, \citenamefont {Omran}, \citenamefont
  {Pohl}, \citenamefont {Gross}, \citenamefont {Kuhr},\ and\ \citenamefont
  {Bloch}}]{Schauss2012}%
  \BibitemOpen
  \bibfield  {author} {\bibinfo {author} {\bibfnamefont {P.}~\bibnamefont
  {Schau{\ss}}}, \bibinfo {author} {\bibfnamefont {M.}~\bibnamefont {Cheneau}},
  \bibinfo {author} {\bibfnamefont {M.}~\bibnamefont {Endres}}, \bibinfo
  {author} {\bibfnamefont {T.}~\bibnamefont {Fukuhara}}, \bibinfo {author}
  {\bibfnamefont {S.}~\bibnamefont {Hild}}, \bibinfo {author} {\bibfnamefont
  {A.}~\bibnamefont {Omran}}, \bibinfo {author} {\bibfnamefont
  {T.}~\bibnamefont {Pohl}}, \bibinfo {author} {\bibfnamefont {C.}~\bibnamefont
  {Gross}}, \bibinfo {author} {\bibfnamefont {S.}~\bibnamefont {Kuhr}}, \ and\
  \bibinfo {author} {\bibfnamefont {I.}~\bibnamefont {Bloch}},\ }\href
  {\doibase 10.1038/nature11596} {\bibfield  {journal} {\bibinfo  {journal}
  {Nature}\ }\textbf {\bibinfo {volume} {491}},\ \bibinfo {pages} {87}
  (\bibinfo {year} {2012})}\BibitemShut {NoStop}%
\bibitem [{\citenamefont {Britton}\ \emph {et~al.}(2012)\citenamefont
  {Britton}, \citenamefont {Sawyer}, \citenamefont {Keith}, \citenamefont
  {Wang}, \citenamefont {Freericks}, \citenamefont {Uys}, \citenamefont
  {Biercuk},\ and\ \citenamefont {Bollinger}}]{Britton2012}%
  \BibitemOpen
  \bibfield  {author} {\bibinfo {author} {\bibfnamefont {J.~W.}\ \bibnamefont
  {Britton}}, \bibinfo {author} {\bibfnamefont {B.~C.}\ \bibnamefont {Sawyer}},
  \bibinfo {author} {\bibfnamefont {A.~C.}\ \bibnamefont {Keith}}, \bibinfo
  {author} {\bibfnamefont {C.-C.~J.}\ \bibnamefont {Wang}}, \bibinfo {author}
  {\bibfnamefont {J.~K.}\ \bibnamefont {Freericks}}, \bibinfo {author}
  {\bibfnamefont {H.}~\bibnamefont {Uys}}, \bibinfo {author} {\bibfnamefont
  {M.~J.}\ \bibnamefont {Biercuk}}, \ and\ \bibinfo {author} {\bibfnamefont
  {J.~J.}\ \bibnamefont {Bollinger}},\ }\href {\doibase 10.1038/nature10981}
  {\bibfield  {journal} {\bibinfo  {journal} {Nature}\ }\textbf {\bibinfo
  {volume} {484}},\ \bibinfo {pages} {489} (\bibinfo {year}
  {2012})}\BibitemShut {NoStop}%
\bibitem [{\citenamefont {Islam}\ \emph {et~al.}(2013)\citenamefont {Islam},
  \citenamefont {Senko}, \citenamefont {Campbell}, \citenamefont {Korenblit},
  \citenamefont {Smith}, \citenamefont {Lee}, \citenamefont {Edwards},
  \citenamefont {Wang}, \citenamefont {Freericks},\ and\ \citenamefont
  {Monroe}}]{Islam2013}%
  \BibitemOpen
  \bibfield  {author} {\bibinfo {author} {\bibfnamefont {R.}~\bibnamefont
  {Islam}}, \bibinfo {author} {\bibfnamefont {C.}~\bibnamefont {Senko}},
  \bibinfo {author} {\bibfnamefont {W.}~\bibnamefont {Campbell}}, \bibinfo
  {author} {\bibfnamefont {S.}~\bibnamefont {Korenblit}}, \bibinfo {author}
  {\bibfnamefont {J.}~\bibnamefont {Smith}}, \bibinfo {author} {\bibfnamefont
  {A.}~\bibnamefont {Lee}}, \bibinfo {author} {\bibfnamefont {E.}~\bibnamefont
  {Edwards}}, \bibinfo {author} {\bibfnamefont {C.-C.}\ \bibnamefont {Wang}},
  \bibinfo {author} {\bibfnamefont {J.}~\bibnamefont {Freericks}}, \ and\
  \bibinfo {author} {\bibfnamefont {C.}~\bibnamefont {Monroe}},\ }\href
  {\doibase 10.1126/science.1232296} {\bibfield  {journal} {\bibinfo  {journal}
  {Science}\ }\textbf {\bibinfo {volume} {340}},\ \bibinfo {pages} {583}
  (\bibinfo {year} {2013})}\BibitemShut {NoStop}%
\bibitem [{\citenamefont {Bohnet}\ \emph {et~al.}(2016)\citenamefont {Bohnet},
  \citenamefont {Sawyer}, \citenamefont {Britton}, \citenamefont {Wall},
  \citenamefont {Rey}, \citenamefont {Foss-Feig},\ and\ \citenamefont
  {Bollinger}}]{Bohnet2016}%
  \BibitemOpen
  \bibfield  {author} {\bibinfo {author} {\bibfnamefont {J.~G.}\ \bibnamefont
  {Bohnet}}, \bibinfo {author} {\bibfnamefont {B.~C.}\ \bibnamefont {Sawyer}},
  \bibinfo {author} {\bibfnamefont {J.~W.}\ \bibnamefont {Britton}}, \bibinfo
  {author} {\bibfnamefont {M.~L.}\ \bibnamefont {Wall}}, \bibinfo {author}
  {\bibfnamefont {A.~M.}\ \bibnamefont {Rey}}, \bibinfo {author} {\bibfnamefont
  {M.}~\bibnamefont {Foss-Feig}}, \ and\ \bibinfo {author} {\bibfnamefont
  {J.~J.}\ \bibnamefont {Bollinger}},\ }\href {\doibase
  10.1126/science.aad9958} {\bibfield  {journal} {\bibinfo  {journal}
  {Science}\ }\textbf {\bibinfo {volume} {352}},\ \bibinfo {pages} {1297}
  (\bibinfo {year} {2016})}\BibitemShut {NoStop}%
\bibitem [{\citenamefont {Koffel}\ \emph {et~al.}(2012)\citenamefont {Koffel},
  \citenamefont {Lewenstein},\ and\ \citenamefont {Tagliacozzo}}]{Koffel2012}%
  \BibitemOpen
  \bibfield  {author} {\bibinfo {author} {\bibfnamefont {T.}~\bibnamefont
  {Koffel}}, \bibinfo {author} {\bibfnamefont {M.}~\bibnamefont {Lewenstein}},
  \ and\ \bibinfo {author} {\bibfnamefont {L.}~\bibnamefont {Tagliacozzo}},\
  }\href {\doibase 10.1103/PhysRevLett.109.267203} {\bibfield  {journal}
  {\bibinfo  {journal} {Phys. Rev. Lett.}\ }\textbf {\bibinfo {volume} {109}},\
  \bibinfo {pages} {267203} (\bibinfo {year} {2012})}\BibitemShut {NoStop}%
\bibitem [{\citenamefont {Sun}(2017)}]{Sun2017}%
  \BibitemOpen
  \bibfield  {author} {\bibinfo {author} {\bibfnamefont {G.}~\bibnamefont
  {Sun}},\ }\href {\doibase 10.1103/PhysRevA.96.043621} {\bibfield  {journal}
  {\bibinfo  {journal} {Phys. Rev. A}\ }\textbf {\bibinfo {volume} {96}},\
  \bibinfo {pages} {043621} (\bibinfo {year} {2017})}\BibitemShut {NoStop}%
\bibitem [{\citenamefont {Horita}\ \emph {et~al.}(2017)\citenamefont {Horita},
  \citenamefont {Suwa},\ and\ \citenamefont {Todo}}]{Horita2017}%
  \BibitemOpen
  \bibfield  {author} {\bibinfo {author} {\bibfnamefont {T.}~\bibnamefont
  {Horita}}, \bibinfo {author} {\bibfnamefont {H.}~\bibnamefont {Suwa}}, \ and\
  \bibinfo {author} {\bibfnamefont {S.}~\bibnamefont {Todo}},\ }\href {\doibase
  10.1103/PhysRevE.95.012143} {\bibfield  {journal} {\bibinfo  {journal} {Phys.
  Rev. E}\ }\textbf {\bibinfo {volume} {95}},\ \bibinfo {pages} {012143}
  (\bibinfo {year} {2017})}\BibitemShut {NoStop}%
\bibitem [{\citenamefont {Vanderstraeten}\ \emph {et~al.}(2018)\citenamefont
  {Vanderstraeten}, \citenamefont {Van~Damme}, \citenamefont {B\"uchler},\ and\
  \citenamefont {Verstraete}}]{Vanderstraeten2018}%
  \BibitemOpen
  \bibfield  {author} {\bibinfo {author} {\bibfnamefont {L.}~\bibnamefont
  {Vanderstraeten}}, \bibinfo {author} {\bibfnamefont {M.}~\bibnamefont
  {Van~Damme}}, \bibinfo {author} {\bibfnamefont {H.~P.}\ \bibnamefont
  {B\"uchler}}, \ and\ \bibinfo {author} {\bibfnamefont {F.}~\bibnamefont
  {Verstraete}},\ }\href {\doibase 10.1103/PhysRevLett.121.090603} {\bibfield
  {journal} {\bibinfo  {journal} {Phys. Rev. Lett.}\ }\textbf {\bibinfo
  {volume} {121}},\ \bibinfo {pages} {090603} (\bibinfo {year}
  {2018})}\BibitemShut {NoStop}%
\bibitem [{\citenamefont {Humeniuk}(2016)}]{Humeniuk2016}%
  \BibitemOpen
  \bibfield  {author} {\bibinfo {author} {\bibfnamefont {S.}~\bibnamefont
  {Humeniuk}},\ }\href {\doibase 10.1103/PhysRevB.93.104412} {\bibfield
  {journal} {\bibinfo  {journal} {Phys. Rev. B}\ }\textbf {\bibinfo {volume}
  {93}},\ \bibinfo {pages} {104412} (\bibinfo {year} {2016})}\BibitemShut
  {NoStop}%
\bibitem [{\citenamefont {Moessner}\ and\ \citenamefont
  {Sondhi}(2001)}]{Moessner2001}%
  \BibitemOpen
  \bibfield  {author} {\bibinfo {author} {\bibfnamefont {R.}~\bibnamefont
  {Moessner}}\ and\ \bibinfo {author} {\bibfnamefont {S.~L.}\ \bibnamefont
  {Sondhi}},\ }\href {\doibase 10.1103/PhysRevB.63.224401} {\bibfield
  {journal} {\bibinfo  {journal} {Phys. Rev. B}\ }\textbf {\bibinfo {volume}
  {63}},\ \bibinfo {pages} {224401} (\bibinfo {year} {2001})}\BibitemShut
  {NoStop}%
\bibitem [{\citenamefont {Isakov}\ and\ \citenamefont
  {Moessner}(2003)}]{Moessner2003}%
  \BibitemOpen
  \bibfield  {author} {\bibinfo {author} {\bibfnamefont {S.~V.}\ \bibnamefont
  {Isakov}}\ and\ \bibinfo {author} {\bibfnamefont {R.}~\bibnamefont
  {Moessner}},\ }\href {\doibase 10.1103/PhysRevB.68.104409} {\bibfield
  {journal} {\bibinfo  {journal} {Phys. Rev. B}\ }\textbf {\bibinfo {volume}
  {68}},\ \bibinfo {pages} {104409} (\bibinfo {year} {2003})}\BibitemShut
  {NoStop}%
\bibitem [{\citenamefont {Powalski}\ \emph {et~al.}(2013)\citenamefont
  {Powalski}, \citenamefont {Coester}, \citenamefont {Moessner},\ and\
  \citenamefont {Schmidt}}]{Powalski2013}%
  \BibitemOpen
  \bibfield  {author} {\bibinfo {author} {\bibfnamefont {M.}~\bibnamefont
  {Powalski}}, \bibinfo {author} {\bibfnamefont {K.}~\bibnamefont {Coester}},
  \bibinfo {author} {\bibfnamefont {R.}~\bibnamefont {Moessner}}, \ and\
  \bibinfo {author} {\bibfnamefont {K.~P.}\ \bibnamefont {Schmidt}},\ }\href
  {\doibase 10.1103/PhysRevB.87.054404} {\bibfield  {journal} {\bibinfo
  {journal} {Phys. Rev. B}\ }\textbf {\bibinfo {volume} {87}},\ \bibinfo
  {pages} {054404} (\bibinfo {year} {2013})}\BibitemShut {NoStop}%
\bibitem [{\citenamefont {Saadatmand}\ \emph {et~al.}(2018)\citenamefont
  {Saadatmand}, \citenamefont {Bartlett},\ and\ \citenamefont
  {McCulloch}}]{Saadatmand2018}%
  \BibitemOpen
  \bibfield  {author} {\bibinfo {author} {\bibfnamefont {S.~N.}\ \bibnamefont
  {Saadatmand}}, \bibinfo {author} {\bibfnamefont {S.~D.}\ \bibnamefont
  {Bartlett}}, \ and\ \bibinfo {author} {\bibfnamefont {I.~P.}\ \bibnamefont
  {McCulloch}},\ }\href {\doibase 10.1103/PhysRevB.97.155116} {\bibfield
  {journal} {\bibinfo  {journal} {Physical Review B}\ }\textbf {\bibinfo
  {volume} {97}},\ \bibinfo {pages} {155116} (\bibinfo {year}
  {2018})}\BibitemShut {NoStop}%
\bibitem [{\citenamefont {McCulloch}()}]{McCulloch_Private}%
  \BibitemOpen
  \bibfield  {author} {\bibinfo {author} {\bibfnamefont {I.~P.}\ \bibnamefont
  {McCulloch}},\ }\href@noop {} {\bibinfo  {journal} {private communication}\
  }\BibitemShut {NoStop}%
\bibitem [{\citenamefont {Coester}\ and\ \citenamefont
  {Schmidt}(2015)}]{Coester2015}%
  \BibitemOpen
\bibfield  {journal} {  }\bibfield  {author} {\bibinfo {author} {\bibfnamefont
  {K.}~\bibnamefont {Coester}}\ and\ \bibinfo {author} {\bibfnamefont {K.~P.}\
  \bibnamefont {Schmidt}},\ }\href {\doibase 10.1103/PhysRevE.92.022118}
  {\bibfield  {journal} {\bibinfo  {journal} {Phys. Rev. E}\ }\textbf {\bibinfo
  {volume} {92}},\ \bibinfo {pages} {022118} (\bibinfo {year}
  {2015})}\BibitemShut {NoStop}%
\bibitem [{\citenamefont {Knetter}\ and\ \citenamefont
  {Uhrig}(2000)}]{Knetter2000}%
  \BibitemOpen
  \bibfield  {author} {\bibinfo {author} {\bibfnamefont {C.}~\bibnamefont
  {Knetter}}\ and\ \bibinfo {author} {\bibfnamefont {G.~S.}\ \bibnamefont
  {Uhrig}},\ }\href {https://doi.org/10.1007/s100510050026} {\bibfield
  {journal} {\bibinfo  {journal} {Eur. Phys. J. B}\ }\textbf {\bibinfo {volume}
  {13}},\ \bibinfo {pages} {209} (\bibinfo {year} {2000})}\BibitemShut
  {NoStop}%
\bibitem [{\citenamefont {Knetter}\ \emph {et~al.}(2003)\citenamefont
  {Knetter}, \citenamefont {Schmidt},\ and\ \citenamefont
  {tz~S~Uhrig}}]{Knetter2003}%
  \BibitemOpen
  \bibfield  {author} {\bibinfo {author} {\bibfnamefont {C.}~\bibnamefont
  {Knetter}}, \bibinfo {author} {\bibfnamefont {K.~P.}\ \bibnamefont
  {Schmidt}}, \ and\ \bibinfo {author} {\bibfnamefont {G.}~\bibnamefont
  {tz~S~Uhrig}},\ }\href {\doibase 10.1088/0305-4470/36/29/302} {\bibfield
  {journal} {\bibinfo  {journal} {Journal of Physics A: Mathematical and
  General}\ }\textbf {\bibinfo {volume} {36}},\ \bibinfo {pages} {7889}
  (\bibinfo {year} {2003})}\BibitemShut {NoStop}%
\bibitem [{\citenamefont {Matsubara}\ and\ \citenamefont
  {Matsuda}(1956)}]{Matsubara1956}%
  \BibitemOpen
  \bibfield  {author} {\bibinfo {author} {\bibfnamefont {T.}~\bibnamefont
  {Matsubara}}\ and\ \bibinfo {author} {\bibfnamefont {H.}~\bibnamefont
  {Matsuda}},\ }\href {\doibase 10.1143/PTP.16.569} {\bibfield  {journal}
  {\bibinfo  {journal} {Prog. Theor. Phys.}\ }\textbf {\bibinfo {volume}
  {16}},\ \bibinfo {pages} {569} (\bibinfo {year} {1956})}\BibitemShut
  {NoStop}%
\bibitem [{\citenamefont {Guttmann}(1989)}]{Guttmann1989}%
  \BibitemOpen
  \bibfield  {author} {\bibinfo {author} {\bibfnamefont {A.~C.}\ \bibnamefont
  {Guttmann}},\ }\href@noop {} {\emph {\bibinfo {title} {Phase Transitions and
  Critical Phenomena}}},\ edited by\ \bibinfo {editor} {\bibfnamefont
  {C.}~\bibnamefont {Domb}}\ and\ \bibinfo {editor} {\bibfnamefont
  {J.}~\bibnamefont {Lebowitz}},\ Vol.~\bibinfo {volume} {13}\ (\bibinfo
  {publisher} {Academic Press},\ \bibinfo {address} {New York},\ \bibinfo
  {year} {1989})\BibitemShut {NoStop}%
\bibitem [{\citenamefont {R\"ochner}\ \emph {et~al.}(2016)\citenamefont
  {R\"ochner}, \citenamefont {Balents},\ and\ \citenamefont
  {Schmidt}}]{Roechner2016}%
  \BibitemOpen
  \bibfield  {author} {\bibinfo {author} {\bibfnamefont {J.}~\bibnamefont
  {R\"ochner}}, \bibinfo {author} {\bibfnamefont {L.}~\bibnamefont {Balents}},
  \ and\ \bibinfo {author} {\bibfnamefont {K.~P.}\ \bibnamefont {Schmidt}},\
  }\href {\doibase 10.1103/PhysRevB.94.201111} {\bibfield  {journal} {\bibinfo
  {journal} {Phys. Rev. B}\ }\textbf {\bibinfo {volume} {94}},\ \bibinfo
  {pages} {201111} (\bibinfo {year} {2016})}\BibitemShut {NoStop}%
\bibitem [{\citenamefont {Takahashi}(1977)}]{Takahashi1997}%
  \BibitemOpen
  \bibfield  {author} {\bibinfo {author} {\bibfnamefont {M.}~\bibnamefont
  {Takahashi}},\ }\href {\doibase 10.1088/0022-3719/10/8/031} {\bibfield
  {journal} {\bibinfo  {journal} {Journal of Physics C: Solid State Physics}\
  }\textbf {\bibinfo {volume} {10}},\ \bibinfo {pages} {1289} (\bibinfo {year}
  {1977})}\BibitemShut {NoStop}%
\bibitem [{\citenamefont {Smerald}\ \emph {et~al.}(2016)\citenamefont
  {Smerald}, \citenamefont {Korshunov},\ and\ \citenamefont
  {Mila}}]{Smerald2016}%
  \BibitemOpen
  \bibfield  {author} {\bibinfo {author} {\bibfnamefont {A.}~\bibnamefont
  {Smerald}}, \bibinfo {author} {\bibfnamefont {S.}~\bibnamefont {Korshunov}},
  \ and\ \bibinfo {author} {\bibfnamefont {F.}~\bibnamefont {Mila}},\ }\href
  {\doibase 10.1103/PhysRevLett.116.197201} {\bibfield  {journal} {\bibinfo
  {journal} {Physical Review Letters}\ }\textbf {\bibinfo {volume} {116}},\
  \bibinfo {pages} {197201} (\bibinfo {year} {2016})}\BibitemShut {NoStop}%
\bibitem [{\citenamefont {Metcalf}(1974)}]{Metcalf1974}%
  \BibitemOpen
  \bibfield  {author} {\bibinfo {author} {\bibfnamefont {B.}~\bibnamefont
  {Metcalf}},\ }\href {\doibase https://doi.org/10.1016/0375-9601(74)90247-3}
  {\bibfield  {journal} {\bibinfo  {journal} {Physics Letters A}\ }\textbf
  {\bibinfo {volume} {46}},\ \bibinfo {pages} {325 } (\bibinfo {year}
  {1974})}\BibitemShut {NoStop}%
\bibitem [{\citenamefont {Korshunov}(2005)}]{Korshunov2005}%
  \BibitemOpen
  \bibfield  {author} {\bibinfo {author} {\bibfnamefont {S.~E.}\ \bibnamefont
  {Korshunov}},\ }\href {\doibase 10.1103/PhysRevB.72.144417} {\bibfield
  {journal} {\bibinfo  {journal} {Physical Review B}\ }\textbf {\bibinfo
  {volume} {72}},\ \bibinfo {pages} {144417} (\bibinfo {year}
  {2005})}\BibitemShut {NoStop}%
\bibitem [{\citenamefont {Blankschtein}\ \emph {et~al.}(1984)\citenamefont
  {Blankschtein}, \citenamefont {Ma}, \citenamefont {Berker}, \citenamefont
  {Grest},\ and\ \citenamefont {Soukoulis}}]{Blankschtein1984}%
  \BibitemOpen
  \bibfield  {author} {\bibinfo {author} {\bibfnamefont {D.}~\bibnamefont
  {Blankschtein}}, \bibinfo {author} {\bibfnamefont {M.}~\bibnamefont {Ma}},
  \bibinfo {author} {\bibfnamefont {A.~N.}\ \bibnamefont {Berker}}, \bibinfo
  {author} {\bibfnamefont {G.~S.}\ \bibnamefont {Grest}}, \ and\ \bibinfo
  {author} {\bibfnamefont {C.~M.}\ \bibnamefont {Soukoulis}},\ }\href {\doibase
  10.1103/PhysRevB.29.5250} {\bibfield  {journal} {\bibinfo  {journal} {Phys.
  Rev. B}\ }\textbf {\bibinfo {volume} {29}},\ \bibinfo {pages} {5250}
  (\bibinfo {year} {1984})}\BibitemShut {NoStop}%
\bibitem [{\citenamefont {Kosterlitz}\ and\ \citenamefont
  {Thouless}(1973)}]{Kosterlitz1973}%
  \BibitemOpen
  \bibfield  {author} {\bibinfo {author} {\bibfnamefont {J.~M.}\ \bibnamefont
  {Kosterlitz}}\ and\ \bibinfo {author} {\bibfnamefont {D.~J.}\ \bibnamefont
  {Thouless}},\ }\href {\doibase 10.1088/0022-3719/6/7/010} {\bibfield
  {journal} {\bibinfo  {journal} {Journal of Physics C: Solid State Physics}\
  }\textbf {\bibinfo {volume} {6}},\ \bibinfo {pages} {1181} (\bibinfo {year}
  {1973})}\BibitemShut {NoStop}%
\bibitem [{\citenamefont {Moessner}\ \emph {et~al.}(2000)\citenamefont
  {Moessner}, \citenamefont {Sondhi},\ and\ \citenamefont
  {Chandra}}]{Moessner2000}%
  \BibitemOpen
  \bibfield  {author} {\bibinfo {author} {\bibfnamefont {R.}~\bibnamefont
  {Moessner}}, \bibinfo {author} {\bibfnamefont {S.~L.}\ \bibnamefont
  {Sondhi}}, \ and\ \bibinfo {author} {\bibfnamefont {P.}~\bibnamefont
  {Chandra}},\ }\href {\doibase 10.1103/PhysRevLett.84.4457} {\bibfield
  {journal} {\bibinfo  {journal} {Phys. Rev. Lett.}\ }\textbf {\bibinfo
  {volume} {84}},\ \bibinfo {pages} {4457} (\bibinfo {year}
  {2000})}\BibitemShut {NoStop}%
\bibitem [{\citenamefont {Wang}\ \emph {et~al.}(2017)\citenamefont {Wang},
  \citenamefont {Qi}, \citenamefont {Chen},\ and\ \citenamefont
  {Meng}}]{Wang2017}%
  \BibitemOpen
  \bibfield  {author} {\bibinfo {author} {\bibfnamefont {Y.-C.}\ \bibnamefont
  {Wang}}, \bibinfo {author} {\bibfnamefont {Y.}~\bibnamefont {Qi}}, \bibinfo
  {author} {\bibfnamefont {S.}~\bibnamefont {Chen}}, \ and\ \bibinfo {author}
  {\bibfnamefont {Z.~Y.}\ \bibnamefont {Meng}},\ }\href {\doibase
  10.1103/PhysRevB.96.115160} {\bibfield  {journal} {\bibinfo  {journal} {Phys.
  Rev. B}\ }\textbf {\bibinfo {volume} {96}},\ \bibinfo {pages} {115160}
  (\bibinfo {year} {2017})}\BibitemShut {NoStop}%
\bibitem [{\citenamefont {Li}\ \emph {et~al.}(2019)\citenamefont {Li},
  \citenamefont {Liao}, \citenamefont {Chen}, \citenamefont {Zen},
  \citenamefont {Sheng}, \citenamefont {Qi}, \citenamefont {Meng},\ and\
  \citenamefont {Li}}]{Li2019a}%
  \BibitemOpen
  \bibfield  {author} {\bibinfo {author} {\bibfnamefont {H.}~\bibnamefont
  {Li}}, \bibinfo {author} {\bibfnamefont {Y.-D.}\ \bibnamefont {Liao}},
  \bibinfo {author} {\bibfnamefont {B.-B.}\ \bibnamefont {Chen}}, \bibinfo
  {author} {\bibfnamefont {X.-T.}\ \bibnamefont {Zen}}, \bibinfo {author}
  {\bibfnamefont {X.-L.}\ \bibnamefont {Sheng}}, \bibinfo {author}
  {\bibfnamefont {Y.}~\bibnamefont {Qi}}, \bibinfo {author} {\bibfnamefont
  {Z.~Y.}\ \bibnamefont {Meng}}, \ and\ \bibinfo {author} {\bibfnamefont
  {W.}~\bibnamefont {Li}},\ }\href {http://de.arxiv.org/abs/1907.08173}
  {\bibfield  {journal} {\bibinfo  {journal} {arXiv:1907.08173}\ } (\bibinfo
  {year} {2019})}\BibitemShut {NoStop}%
\bibitem [{\citenamefont {Smerald}\ and\ \citenamefont
  {Mila}(2018)}]{Smerald2018}%
  \BibitemOpen
  \bibfield  {author} {\bibinfo {author} {\bibfnamefont {A.}~\bibnamefont
  {Smerald}}\ and\ \bibinfo {author} {\bibfnamefont {F.}~\bibnamefont {Mila}},\
  }\href {\doibase 10.21468/SciPostPhys.5.3.030} {\bibfield  {journal}
  {\bibinfo  {journal} {SciPost}\ }\textbf {\bibinfo {volume} {5}},\ \bibinfo
  {pages} {030} (\bibinfo {year} {2018})}\BibitemShut {NoStop}%
\end{thebibliography}%

\end{document}